\documentclass[twocolumn,showpacs,preprintnumbers,amsmath,amssymb]{revtex4}

\usepackage{graphicx}
\usepackage{dcolumn}
\usepackage{bm}

\usepackage{ulem} %% for strike-through
\usepackage[usenames]{color}

\begin{document}

\preprint{YITP-08-8}

\title{Sigma meson in pole-dominated QCD sum rules}

\author{Toru Kojo$^{1}$ and Daisuke Jido$^2$}

\affiliation{
$^1$Department of Physics, Kyoto University, Kyoto 606-8502, Japan \\
$^2$Yukawa Institute for Theoretical Physics, Kyoto University,
          Kyoto 606-8502, Japan
}

\date{\today}

\begin{abstract}
The properties of $\sigma$(600) meson
are studied using the QCD sum rules (QSR) for the tetraquark operators.
In the SU(3) chiral limit,
we investigate separately
SU(3) singlet and octet tetraquark states
as constituents of the $\sigma$ meson,
and discuss their roles
for the classification of the light scalar nonets, $\sigma, f_0, a_0$, and
$\kappa$, as candidates of tetraquark states.
All our analyses are performed
in the the suitable Borel window which 
is indispensable to avoid the {\it pseudo peak} artifacts
outside of the Borel window.
The acceptably wide Borel window originates
after preparing the favorable set up of a 
linear combination of operators
and the inclusion of the dimension 12 terms in the OPE.
Taking into account for the possible large width,
we estimate masses for singlet and octet states as 
$700\sim 850$ MeV, $600\sim 750$ MeV, respectively, although
octet states have smaller overlap with the pole than singlet state
and may be strongly affected by low energy scattering states.  
This splitting of singlet and octet states 
emerges from the number of the $\bar{q}q$ annihilation diagrams,
which include both color singlet annihilation processes,
$qq\bar{q}\bar{q}\rightarrow (q\bar{q})_1$ and
color octet annihilation processes,
$qq\bar{q}\bar{q}\rightarrow G (q\bar{q})_8$.
The mass evaluation for the $\sigma$ meson
gives the value around $600\sim800$ MeV
which is much smaller than the mass obtained by 2-quark correlators,
$1.0\sim1.2$ GeV. 
This indicates $\sigma$ state has the large overlap
with the tetraquark states.
\end{abstract}


\maketitle

\section{Introduction}
The structure of scalar mesons is a long-standing
problem in hadron spectroscopy \cite{review}.
In contrast to the other hadrons, two flavor nonets appear around 1 GeV 
in the scalar meson spectra.
The systematic classification of scalar mesons
sheds light on the nonperturbative aspects of QCD.
Especially, the lighter scalar nonet, 
the isoscalar $\sigma$(600), $f_0$(980), the isodoublet $\kappa$(800), 
and the isovector $a_0$(980), are candidates for exotic hadrons (exotics)
which have more complicated components 
than simple mesonic $q\bar q$ or baryonic $qqq$ structure.
The naive $q\bar{q}$ assignment of the constituent quark picture
for these mesons,
$\sigma \sim (u\bar{u}+d\bar{d})$, 
$f_0 \sim s \bar{s}$,
$\kappa \sim u\bar{s}$,
$a_0 \sim (u\bar{u}-d\bar{d})$, 
may lead the unrealistic
mass ordering, 
$m(\sigma) \sim m(a_0) < m(\kappa) < m(f_0)$,
and suggest heavier masses due to the P-wave angular excitation between two quarks, 
which typically costs $500$ MeV, for the
$J^{P}=0^{+}$ quantum number of the scalar mesons.
It is believed that these $q \bar q$ assignments
are most likely realized in the nonet
above 1 GeV rather than the nonet below 1 GeV. 

The masses of the light scalar mesons are explained 
by several pictures. One of them is the four quark picture 
proposed by Jaffe \cite{Jaffe}.
In this picture, assuming the quark contents with the ideal
mixing of the flavor as 
$\sigma \sim (ud)(\bar{u}\bar{d})$, 
$f_0 \sim [(ds)(\bar{d}\bar{s}) + (su)(\bar{s}\bar{u})]$,
$\kappa \sim (ud)(\bar{s}\bar{u})$,
$a_0 \sim [(ds)(\bar{d}\bar{s})-(su)(\bar{s}\bar{u})]$,
one can naturally explain the observed mass ordering.
For $\sigma$ meson, 
quenched lattice calculations also support
this assignment \cite{lattice, Kentucky}.
Qualitatively, one of the key ideas for the considerable mass reduction 
of the light scalar mesons below 1~GeV
is to employ the possible strong diquark correlation
originated from the chromo-magnetic interaction~\cite{Jaffe}.
This brings us to an
interesting possibility that the scalar mesons
are considered as the good place to investigate the strength of diquark correlation,
which may provide a useful building block to understand
the hadron spectra \cite{Jaffe2} and for the further applications
to the hot/dense quark matter \cite{quark matter}.

Mixing of the two and four quark components is also employed to
explain the scalar meson spectra. For example, the scalar nonet above 1 GeV
have larger mass than one expected from the $q\bar{q}$ picture in the
conventional quark models, as seen
in the anomalous spin-orbit splitting, $a_1(1230)<a_2(1320)<a_0(1450)$,
in contrast to the charmonium mass splitting
$\chi_{c0}(3414)<\chi_{c1}(3510)<\chi_{c2}(3556)$.
This mass ordering could be explained by a mixing scenario 
of the two and four quark components, in which, as a result of the level repulsion,
the mass of the four-quark dominate state gets reduced, while the two-quark
dominated state is pushed up \cite{Higgs}.

For the isoscalar sector, $\sigma$ and $f_0$,
Narison discussed another possibility \cite{Narison} 
invoking the QCD sum rule (QSR) \cite{Shifman, Reinders} and some low energy theorems \cite{LET}.
In his scenario,
the glueball degrees of freedom 
come into play and
the strong $q\bar{q}-G^2$ mixing leads the considerable
mass reduction below 1 GeV
although unmixed glueball and $q\bar{q}$ are expected to be
relatively heavy, $\sim 1.6$ GeV \cite{Glueball} 
and $\sim1.4$ GeV \cite{Kentucky}, respectively,
in the quenched lattice calculations.
The strength of the  $q\bar{q}-G^2$ mixing has been studied in various 
approaches, but the strong mixing has not been confirmed yet \cite{mixing}.

The hadronic picture for the scalar meson were also widely discussed.
The $\sigma$, $\kappa(800)$, $f_{0}(980)$ and $a_{0}(980)$ are
dynamically generated as quasi-bound resonance states in scattering of 
the Nambu-Goldstone bosons based on chiral dynamics with an appropriate 
treatment for restoring unitarity in the scattering amplitudes \cite{chiral}. 

The $\sigma$ meson itself is an attractive subject of contemporary nuclear
physic.
The $\sigma$ field is the origin of the attractive part of the nuclear force 
in the intermediate energy region \cite{nuclear}
and can be a chiral partner of the pion being a possible soft mode
in the chiral restoration \cite{Kunihiro}.
Its existence of the $\sigma$ meson had 
been a long-standing problem,
but recent experemental results
such as $D^+ \rightarrow \pi^+ \pi^+ \pi^-$ \cite{E791},
$J/\psi \rightarrow \omega \pi^+ \pi^-$ \cite{BES},
and dispersion analysis employing Roy equation \cite{Roy}
support the existence of the $\sigma$ pole
with mass $440\sim 540\ {\rm MeV}$ and width
$250\sim 540\ {\rm MeV}$.

Yet the structure of the scalar mesons is not conclusive. We have 
several  pictures of the constituents of the scalar meson with their
admixtures. 
One of the purposes in this paper is
to discuss the relevant constituents and the mixing scheme.
They can be investigated through comparison 
of the several correlators utilizing in the QSR \cite{Shifman, Reinders},
which relates the nonperturbative aspects of QCD 
to the hadronic properties.
In this work, our main consideration is on the
four-quark (4q) picture for the light scalar nonet
and we employ the tetraquark operators to
obtain large overlap with the tetraquark states.
We will show that the correlators of the tetraquark operators 
include not only  four-quark connected diagrams but also disconnected 
diagrams, such as the
$(q\bar{q})_1$ and $(q\bar{q})_8 G$ in the intermediate states.
These mixing effects can be studied
from the difference between SU(3) flavor
singlet and octet correlators.
In fact, the number of the annihilation diagrams
is four times larger in the singlet case than in 
the octet case.
In our operator case,
this leads larger low energy enhancement in 
the flavor singlet case.
Throughout this paper,
to avoid the complication from the current quark mass effects,
we take the SU(3) chiral limit, in which 
the singlet and octet states completely decouple.

For the tetraquark analyses with QSR of the $\sigma$ meson, 
we have several technical issues to be solved. The most important
point is that the resonance mass should be extracted within so-called
{\it Borel window} in which the QSR works with better accuracy. 
The Borel mass $M$ is introduced to the QSR by a derivative operation 
$L_{M}$ to the dispersion relation of the correlation function $\Pi$ as
\begin{eqnarray} 
L_M \Pi^{ope}(-Q^2) = \int_0^{\infty} \!\! ds\ e^{-s/M^2}\frac{1}{\pi} {\rm Im}\Pi^{h}(s),
\label{eq:qsrdisp}
\end{eqnarray}
where $L_{M}\equiv d^{n}/(dQ^{2})^{n}$ taking $n\rightarrow \infty$ 
and $Q^{2} \rightarrow \infty$ being fixed by $Q^{2}/n = M^{2}$.
The correlation function
$\Pi^{ope}$ in the left hand side is calculated by the operator product expansion (OPE)
and $\Pi^{h}$ in the right hand side is expressed by hadronic contributions.
The tower of successes in reproducing the meson and baryon spectra
is attributed to appropriate and careful applications of QSR 
in suitable Borel mass region, i.e., {\it Borel window}.
Only within the Borel window, one is allowed to extract the low energy properties 
of the spectral integral. 

The main technical problem for the multi-quark QSR is that
the setting up the Borel window becomes quite difficult.
In contrast to the
usual meson and baryon cases \cite{KHJ}, 
since correlation functions of multi-quark interpolating fields
show slow convergence of OPE and unwelcome high energy contamination 
dominates the spectral integrals. 
As we will emphasize in Sec.\ref{sec:Basics} and \ref{sec:artifact}, 
this leads difficulties to extract low energy properties 
of the spectral function and we are often stuck with
the {\it pseudo} peak artifacts outside of the Borel window.

All these issues can be solved by use of suitable interpolating fields
and inclusion of the OPE terms up to dimension 12 (dim.12)
for the tetraquark operators. After that, the Borel windows are found and
we can investigate the physical quantities within the Borel windows. 
To complete solving these issues, we would like to add one more analysis
for the $\sigma$ meson in QSR with the tetraquark operator despite 
tetraquark operator analysis for the $\sigma$ meson, although there have
been many works done in the past \cite{noBW, rcnp}. 
 
Another important aspect when we discuss 
the $\sigma$ meson is its
possible large width and the
effects on the mass evaluation of $\sigma$.
In Sec.\ref{sec:width}, 
we discuss the Borel transformed Breit-Wigner type spectral function
and test how they behave in the effective mass plot.
We find the stablity is moderate even up to $\sim 400\ {\rm MeV}$ width.
As related to this, we re-examine the usual criteria to fix $s_{th}$.
They are reflected to the estimations for
the physical quantities of the $\sigma$ meson.

The organization of this paper is as follows.
In Sec.II, we explain the basic concepts of QSR and
illustrate the typical pseudo-peak artifacts
which we often encounter.
The importance to set the Borel window is emphasized.
The effects of the width on the effective mass plot
are illustrated, and
the criterion to fix $s_{th}$ is re-examined.
In Sec.III, we discuss the flavor singlet and octet states in the chiral limit.
We also argue that both color singlet and octet annihilation diagrams
are source to split the singlet and octet states.
This splitting may be important
to classify $f_0$, $\kappa$, $a_0$ even after taking realistic quark mass,
which will be discussed in the subsequent paper.
In Sec.IV, we show the results of the Borel analyses for the sigma meson 
together with the singlet and octet states. At first we introduce
the tetraquark operator used in the analyses as a linear combination 
of two types of diquark-diquark local operators. We discuss the criteria
to determine the mixing angle of the operators.
After fixing it, we perform the Borel analysis. 
For the singlet state
we obtain the mass value $700 \sim 850\ {\rm MeV}$, 
while, for octet state, we estimate the mass $600 \sim 750\ {\rm MeV}$,
although the residue of the octet state is much smaller than the singlet
state and thus may be affected by low energy scattering background.
Finally we examine the $\sigma$ meson 
as superposition of singlet and octet states
and estimate its mass $600\sim 800\ {\rm MeV}$ within error
of width effects.
The Sec.V and VI are devoted to discussion and summary, respectively.
All the details about the OPE terms and operator dependence
are summarized in the Appendix.

\section{The basic concepts of QSR
and possible pseudo-peak artifacts }
\label{sec:basicsconcept}

In this section, we start with a brief review of 
the basic concepts of QSR,
especially emphasizing the importance of the Borel window.
Definitions of the terminologies and notations
used in the later analyses are given in Sec.\ref{sec:Basics}.
In Sec.\ref{sec:artifact}, we illustrate a 
typical artifact in the sum rules,
{\it pseudo peak artifact}, which often appears in QSR for the
exotic hadrons.
We show that imposing correct criteria on the Borel window rejects 
such an artifact. 
In Sec.\ref{sec:width}, we examine the width effects on 
the Borel stability plots for the effective mass and residue.
The way to determine the threshold parameter $s_{th}$ is also
discussed in detail.

\subsection{Basic concepts of QSR}
\label{sec:Basics}
Following the standard way of the QCD sum rule, 
we start with the time-ordered two-point correlation function
of a tetraquark interpolating field $J$ for the scalar meson:
\begin{eqnarray}
 \Pi (q^{2})
\equiv i\int\! d^{4}x\ e^{iq\cdot x} \langle 0 | T[J(x) J^\dag(0)] | 0 \rangle,
\label{eq:corrfunc}
\end{eqnarray}
where $\langle 0|\cdots|0 \rangle$ denotes a vacuum expectation value.
(Hereafter we write it as $\langle \cdots\rangle$ for brevity.)
The QSR is then obtained 
through the dispersion relation,
\begin{equation}
{\rm Re}\Pi(q^2) = {\rm P} \int_0^{\infty} \! ds\,
\frac{1}{\pi} \frac{ {\rm Im}\Pi(s) } {(s-q^2)}  
\label{eq:disp}
\end{equation}
satisfying the the spectral conditions, ${\rm Im}\Pi(s) \geq 0 $.
For sufficiently large $-q^2$, 
the left hand side of (\ref{eq:disp}) can be expressed by the 
operator product expansion (OPE) with $C_i$ including the vacuum condensates:
\begin{eqnarray}
\Pi_i^{ope}(q^2)
= \sum_{j=0}^{4}C_{2j}\ (q^2)^{4-j}\log(-q^2)
+ \sum_{j=1}^\infty\frac{C_{8+2j}}{(q^2)^j}.\,
\label{eq:ope}
\end{eqnarray}
The OPE starts from $(q^{2})^{4} \log (-q^{2})$ 
reflecting the large number of quark fields
in the tetraquark interpolating field. 
This turns out to be the main origin of 
QSR artifacts as discussed in Sec.\ref{sec:artifact}.

The imaginary part in the right hand side of Eq.(\ref{eq:disp}) 
is deemed to be the hadronic spectrum. 
Based on the quark hadron duality ansatz,
we approximate higher energy part of the spectral function
than a threshold $s_{th}$ to 
the spectral function obtained by OPE:
\begin{eqnarray}
{\rm Im} \Pi^{h}(s) = \theta(s_{th}-s)\,{\rm Im}\Pi^h(s)
+ \theta(s-s_{th})\,{\rm Im}\Pi^{ope}(s). \label{eq:divide}
\end{eqnarray}
Here we introduce the threshold parameter $s_{th}$
where the quark hadron duality ansatz begins to work.
(We use the notation $E_{th}\equiv \sqrt{s_{th}}$ in the following.)
Hereafter, we write the first term of the right hand side of Eq.\eqref{eq:divide} as
\begin{eqnarray}
\Pi^<(s) \equiv
\theta(s_{th}-s)\,{\rm Im}\Pi^h(s)
\end{eqnarray}
for the later convenience.
In usual QSRs, the low energy part
$\Pi^{<}$ is parametrized by a delta function,
$\lambda^2 \delta(s-m^2)$
with the pole mass $m$ and the overlap residue $\lambda$ 
of the interpolating field $J$ and the hadronic state.
Instead, we do not specify the form of $\Pi^{<}$ at this stage, since we also
consider more general contributions from the resonance width and background 
scattering.

Substituting the expressions Eqs.(\ref{eq:ope}) and (\ref{eq:divide})
into Eq.(\ref{eq:disp}) and using Eq.(\ref{eq:qsrdisp}), we obtain the Borel transformed
expression of the sum rule
\begin{eqnarray}
\int_0^{s_{th}}\! ds \ e^{-s/M^2} 
\frac{1}{\pi} {\rm Im}\Pi^<(s)
= \sum_{j=1}^{\infty}
 \frac{ (-)^j }{ \Gamma(j) }
\frac{ C_{ 8+2j } }{ (M^2)^{j-1} } && \nonumber \\
+ \bigg( \int_0^\infty - \int_{s_{th}}^\infty \bigg)\ ds
\ e^{-s/M^2} \sum_{j=0}^{4}C_{2j}\ s^{4-j}, &&
\label{sum rule2}
\end{eqnarray}
with the Gamma function $\Gamma(n)=(n-1) !$.
Calculating the Wilson coefficients in the right side, 
we can evaluate the hadronic parameters in the left side as outputs.
Here the Borel transformation impoves the OPE convergence
by factor $1/\Gamma(j)$. 
At the same time, this transformation
reduces the contaminations from high energy states
due to the exponential factor $e^{-s/M^2}$.

Using the equality (\ref{sum rule2}), the {\it effective} mass and residue 
are derived as
\begin{eqnarray}
m_{ {\rm eff} }^2(M^2) \equiv \frac{ \int_0^{s_{th}} \!ds\ e^{-s/M^2}\ s\ {\rm Im} \Pi^< (s) }
 { \int_0^{s_{th}} \!ds\ e^{-s/M^2}\ {\rm Im} \Pi^< (s) },
\hspace{1.0cm} && \label{mass} \\
 \lambda_{ {\rm eff} }^2(M^2) \equiv e^{m_{ {\rm eff} }^2(M^2)/M^2} \times  \int_0^{s_{th}} \!ds\
  e^{-s/M^2}\ {\rm Im} \Pi^< (s), &&
\label{residue}
\end{eqnarray}
in the similar way to mass evaluation in lattice calculations.
The reason that we call these values as ``effective'' mass and residue is
that $m_{ {\rm eff} }$ and $\lambda_{ {\rm eff} }$ include
not only pole mass contribution but also 
the effects of the width and the background.
In the usual one peak ansatz, 
the width and background effects
are regarded as small, and then 
the physical variables should behave independently on the Borel mass
and $s_{th}$ is chosen to satisfy this criterion.
(More detailed discussion for how to fix $s_{th}$ is
given in Sec.\ref{sec:width}.)

Now we turn to the explanation for
the situation where QSR is workable.
In any estimation of the physical quantities,
it is required that Eq.(\ref{sum rule2}) holds with good accuracy.
This is satisfied after the suitable selection of the Borel mass
which achieves the good OPE convergence 
and reduces the unwanted contaminations from
high energy states.
Such a region for the Borel mass is called Borel window.
This is important 
both conceptually and practically to derive
definite results.
A conceptual importance is already emphasized just above.
Practically, the Borel window is useful to reject
the artifacts.
An explicit example will be shown
in Sec.\ref{sec:artifact}.

The Borel window in our analysis is determined as follows based on
Ref.~\cite{Reinders}:
The lower boundary of the window is set up so as to make
the OPE convergence sufficient in higher-dimensional operators.
The criterion is
quantified so that the highest-dimensional terms in the truncated OPE are
less than 10\% of its whole OPE, i.e.,
\begin{eqnarray}
A(M^2) \equiv \bigg| 
\frac{ {\rm dim.{\it n}\ terms} }{ {\rm OPE\ summed\ up\ to\
 dim.{\it n} } } \bigg|
\le 0.1.
\label{opeconv}
\end{eqnarray}
At the same time, the higher boundary
of the window is fixed by the pole-dominance condition
that
\begin{eqnarray}
B(M^2;s_{th}) \equiv 
\frac{ \int_0^{ s_{th} } ds\ e^{-s/M^2} {\rm Im}\Pi^<(s) }
     { \int_0^{\infty} ds\ e^{-s/M^2} {\rm Im}\Pi(s) } \ge 0.5. 
\label{poleratio}
\end{eqnarray}

\begin{figure*}[floatfix]
  \begin{center}
    \begin{tabular}{cc}
      \resizebox{70mm}{!}{\includegraphics{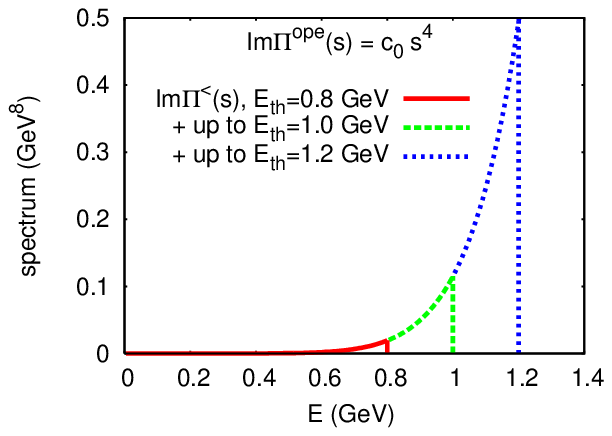}} &
      \resizebox{70mm}{!}{\includegraphics{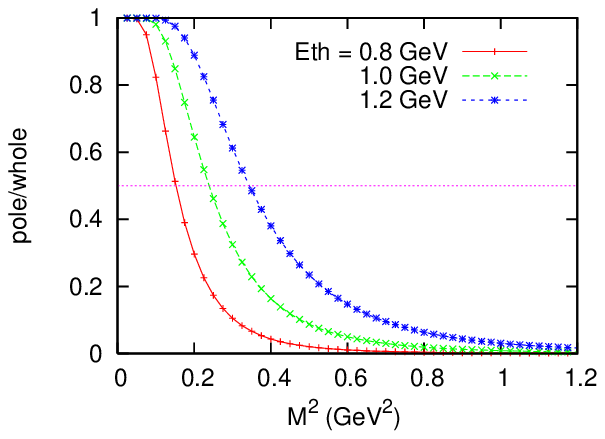}} \\
      \resizebox{70mm}{!}{\includegraphics{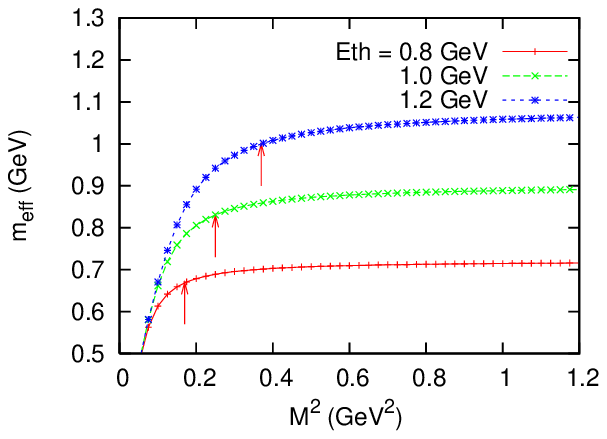}} &
      \resizebox{70mm}{!}{\includegraphics{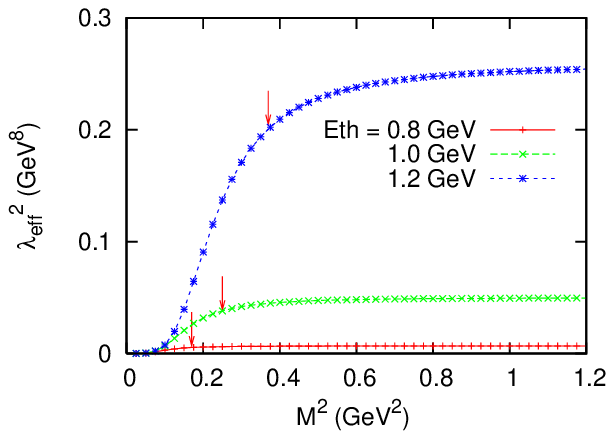}} \\
    \end{tabular}
    \caption{(Color online) The examples of the pseudo peak artifacts.
   The upper left panel is the plot of the
   spectral function with the dim.0 term of the OPE
   up to the thresholds, $E_{th}=0.8$, $1.0$ and $1.2$ GeV.
   The rapid enhancement of the 
   polynomial $s^4$ mimics the peak just below $s_{th}=E_{th}^2$. 
   The upper right panel shows the rate
   of the pole dominance,
   $B(M^2;s_{th})$.
   Without higher dimension terms, the rate is very small
   because of the lack of the low energy correlation.
   The lower panels show the mass in the left panel and residue
   in the right panel, respectively.
   The arrows indicate the $M^2$ values where the pole dominance 
   ratio begins to fall below 50\%, 
   which is the upper bound of the Borel window. 
   In the larger $M^2$, both quantities show the moderate stability
   because of the pseudo peak behavior just below $s_{th}$.}
    \label{artifact}
   \vspace{-0.5cm}
  \end{center}
\end{figure*}

Setting up the Borel window is the most important
step for the application of the sum rules to discuss 
the low energy side in the spectral integral,
especially for exotic hadrons, as emphasized in \cite{KHJ}.
However, this important step was sometimes neglected
in the literature for the exotics.
Therefore their successes seem to be partial.
Outside of the Borel window, 
we are stuck with the sum rules artifacts,
i.e., the artificial stability of the physical quantities
against the variation of the Borel mass.
Further, the evaluation of the 
physical properties, mass, residue and so on,
depends on the selection of $s_{th}$ so strongly
that QSR loses the predictive power. 
In the next subsection, we will illustrate
the origin of the artifacts using some examples. 

%%%%%%%%%%%%%%%%%%%%%%%%%%%%%%%%%%%%%%%%%%%%%%%%%%%%%%%%%%%%%%%
\subsection{Pseudo peak artifacts}
\label{sec:artifact}
\begin{figure*}[floatfix]
  \begin{center}
    \begin{tabular}{ccc}
      \hspace{-0.5cm} \resizebox{60mm}{!}{\includegraphics{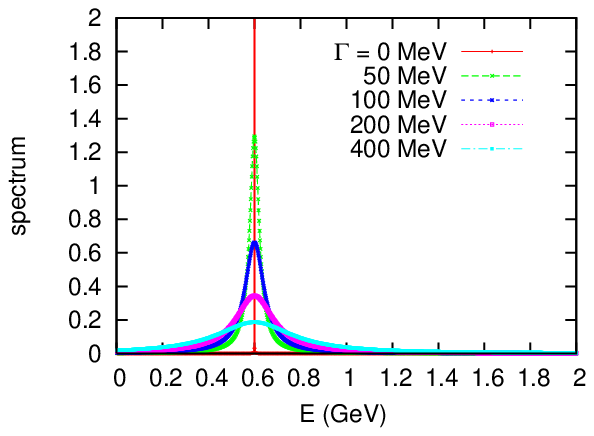}} &
      \hspace{-0.5cm} \resizebox{60mm}{!}{\includegraphics{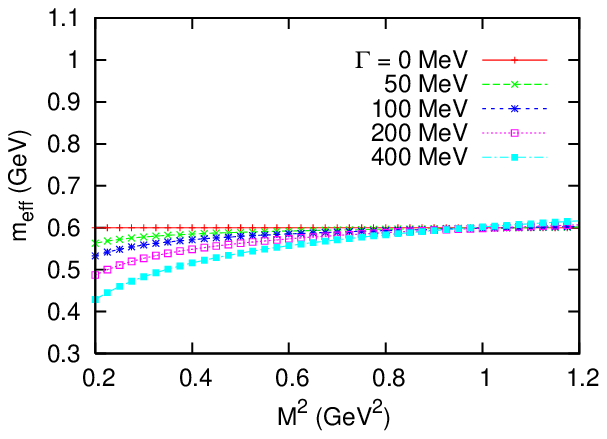}} &
      \hspace{-0.5cm} \resizebox{60mm}{!}{\includegraphics{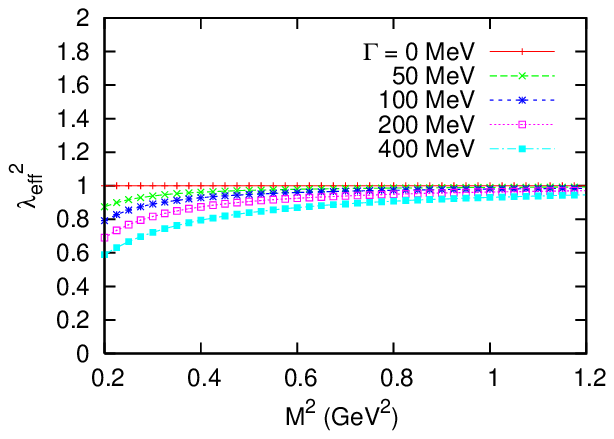}} \\
    \end{tabular}
    \caption{(Color online) The relations among the Breit-Wigner form of the
   spectral function, the effective mass and the residue plots.
   The left panel shows the Breit-Wigner spectral function with the fixed mass 600 MeV
   and the values of the width are taken
   as 0, 50, 100, 200 and 400 MeV.
   The middle and right figures show the effective mass and 
   residue calculated by Eqs.\eqref{mass} and \eqref{residue} with the Breit-Wigner
   spectral function, respectively.
   Because of the low energy tails, the effective mass and residue plots
   shift downward in the case of the small $M^2$.}
   \label{breit}
  \end{center}
\end{figure*}

In the following arguments, we consider one of the simplest  
spectral functions
having only the dim.0 term 
proportional to $s^4$ in the tetraquark operator case.
This case study provides a good example 
to examine typical artifacts in the QSR for the exotics. 
Similarly this is the case if we consider only
the lower dimension terms
proportional to $s^n\ (n>0)$,
and then we recognize
the importance of the higher dimension terms dim.8, 10, 12,...
beyond the polynomials in OPE.

Equation (\ref{sum rule2}) with only dim.0 term becomes
\begin{eqnarray}
\int_0^{s_{th}}\!\! ds \ e^{-s/M^2} 
\frac{1}{\pi} {\rm Im}\Pi^<(s)
= \int_0^{s_{th}}\!\! ds
\ e^{-s/M^2} C_0 s^4,
\label{sum rule3}
\end{eqnarray}
where the value of the coefficient $C_0$ is irrelevant in the following discussion.
We expect that the dim.0 term poorly 
contains the low energy correlation,
because the polinomial $s^4$ drastically decreases in small $s$ region, 
while $s^4$ increases rapidly with larger $s$.
Then if we use such a simply increasing function as the spectral function for the sum rules, 
we encounter unphysical Borel stability for the mass and residue, 
since the function cut above $E_{th}$ behaves like a peak function 
as seen in the upper left panel in Fig.\ref{artifact}.
We call this {\it pseudo peak} artfact which 
often appears in the QSR without including enough
higher dimension terms
beyond the simple polynaomial $s^n\ (n>0)$.

Let us see how {\it pseudo peak} artifact affects the QCD sum rules.
As seen in the lower panels of Fig.\ref{artifact},
both effective mass and residue 
plots show the stability in the larger $M^2$ region,
as a result that the spectral function just below
$E_{th}$ behaves like a peak without large suppression
from the factor $e^{-s/M^2}$.
It is important point that
they exhibit the strong sensitivity to the
threshold parameter.
This is a consequence that the $E_{th}$ value 
directly determines the position of the {\it pseudo peak}.
For the effective mass illustrated in lower left panel of Fig.\ref{artifact},
the change of $E_{th}$ by 200 MeV
leads the mass change of typically $150$ MeV.
The strong change is unphysical
because $E_{th}$, in principle,
does not have any direct relations with the position of the resonance peak.
The threshold dependence is more clearly seen in the effective residue plot.
Since the spectrum increases like $s^4$,
taking larger $E_{th}$ leads the drastic increase of the {\it pseudo peak}
strength, as illustrated in lower right panel of Fig.\ref{artifact}.

This artifact can be easily found out 
by the examination of the 
rate of the pole dominance even when we treat 
more complicated spectral functions than that we use now.
Shown in the upper right panel of Fig.\ref{artifact} is 
the pole dominance ratio $B(M^2;s_{th})$ defined in Eq.\eqref{poleratio} as the function of $M^2$,
which is typical quantity to measure
the strength of the low energy correlation
compared to the unwanted high energy correlation.
The exponential factor in Eq.(\ref{sum rule3})
enables to extract the low energy part of the spectral function.
The ratio $B(M^{2};s_{th})$ determines the upper bound of the
applicable range of the sum rule in terms of the Borel mass (Borel window).
As seen in two lower panels of Fig.\ref{artifact},
the artificial stability is seen above the upper bound of the Borel
window, which is indicated by arrow in the figure.
In this way, the condition of the sufficient pole dominance
rejects the results with artificial stabilities.
This is one of the practical usages of the Borel window.

Throughout this section,
we have seen the origin of the QSR artifact
outside of the Borel window.
The lessons we can learn from these examples 
are the terms with higher dimension like dim.8
have crucial importance to include
low energy correlation and to be free from artifacts.
In connection with the criterion on the Borel window,
the inclusion of higher dimension terms improves
the pole dominance and considerably extend the upper bound of the
Borel window.
In addition to the ensurance of the OPE convergence,
this is another reason that 
we include the higher dimension terms up to dim.12.
%
%%%%%%%%%%%%%%%%%%%%%%%%%%%%%%%%%%%%%%%%%%%%%%%%%%%%%%%

\subsection{Possible width effects on the plot of the physical quantites
and examination for the threshold fixing criterion}
\label{sec:width}

Most of the light scalar nonets
are considered to have large widths typically $100 \sim 400$ MeV
though they have not been well-determined yet.
It is important to check how their widths
affect the effective mass and residue plots.
In this section, we employ as a test function 
a simplest spectral function including the width,
i.e., Breit-Wigner function (BWF),
\begin{eqnarray}
{\rm const.} \times \frac{\Gamma^2/4}{(\sqrt{s}-m)^2 + \Gamma^2/4},
\end{eqnarray}
where {\it m} is the pole mass and $\Gamma$ is the width.
The {\rm const.} will be taken to normalize the integral 
of this function to 1.
We show in the first panel of Fig.\ref{breit}
the BWF with the pole mass fixed to 600 MeV and
the width taken as 0, 50, 100, 200, and 400 MeV.

Let us see the behavior of the physical quantities as
a consequence after the substitution BWF 
into ${\rm Im}\Pi^<(s)$ of the Eqs.(\ref{mass}) and (\ref{residue}).
The middle and right panels in Fig.\ref{breit} show
the effective mass and residue plots, respectively.
In the case of the zero width approximation,
both mass and residue plots show the 
complete independence on the variation of $M^2$
as expected.
On the other hand, in the nonzero width cases,
BWF has the tail both in the lower and higher energy
around the pole mass, and then
the effective mass shifts downward (upperward) in the lower (higher) $M^2$.

Fig.\ref{breit} tells us that, if we consider the effect 
of the width in the effective mass plot, the best Borel stability does
not necessarily give the best fit of the spectral function. 
This violation of the Borel stability due to the width
affects how to select the $E_{th}$ and the final results.
In the zero width or very narrow width case,
usually we can fix $E_{th}$ to give the best Borel stability because 
the effective mass plots of the zero width case
do not depend on the Borel mass $M^{2}$.
But in the case with a wider width, how should we fix $E_{th}$?

\begin{figure}[t]
 \resizebox{80mm}{!}{ \includegraphics{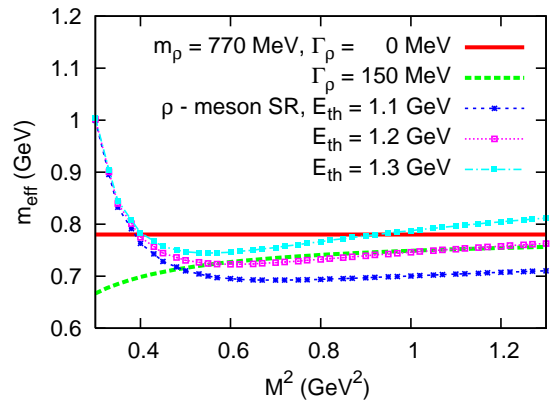} }
\caption{(Color online) The effective mass for the Breit-Wigner function
with $m_{\rho}$=770 MeV, $\Gamma_{\rho}$=0, 150 MeV,
and for the $\rho$-meson SR 
results with $E_{th}$=1.1, 1.2, 1.3 GeV. }
\label{massrho}
\end{figure}

If we assume the shape of the spectrum for the resonances,
the effective mass plots obtained by QSR can be compared with those of the known spectrum,
and we can choose $E_{th}$ to achieve the better fit.   
This can be seen in the following example.
Let us consider the $\rho$ meson which has the mass $m_{\rho}=770$ MeV 
and the width $\Gamma_{\rho}=150$ MeV,
and assume that the spectral function is given by the
Breit-Wigner form.
We show in Fig.\ref{massrho} the
effective mass plots for the BWF with $m_{\rho}=770$ MeV and
$\Gamma_{\rho}=0,\ 150$ MeV, and 
also the effective mass plot obtained by the QSR with the OPE given
in the literature \footnote{Here the OPE
is calculated up to dim.6 and the violation of 
the condensate factorization is considered with factor 2.
The values of the condensates is the same as those used in our analyses in this
paper, summarized in Sec.\ref{SecBorel}. }
with several thresholds $E_{th}=1.1,\ 1.2,\ 1.3$ GeV.
In the zero width approximation,
$E_{th}=1.1$ GeV is prefered within QSR analyses 
because the Borel stability is achieved in the effective mass plot best.
In this case, the mass is read as 0.7 GeV from the Borel stability.
Once the effects of the width is considered in the effective mass plot,
we lose the reason to select $E_{th}$ which gives the best stability. 
But, comparing the effective mass plot of the BWF with the 150 MeV width,
we find that the case of $E_{th}=1.2$ GeV, 
in which the stability is slightly 
worse than the $E_{th}=1.1$ GeV case, is better 
to reproduce the effective mass plot for BWF with 
the 770 MeV mass and the 150 MeV width.
In this way, 
we can estimate the threshold $E_{th}$ and the resonance mass $m$.
Nevertheless, for the scalar meson case, we do not know neither the shape of the
spectral function nor the width of the resonance, {\it a priori}. 
Thus, we cannot definitely 
obtain the exact values of the threshold and the resonance mass. In this analysis,
the resonance masses are estimated by assuming the zero width and
the 400 MeV width with the BWF. Then we adopt the range between two masses
as our result of QSR. As we will see below, the difference of these two mass
is about 20\%.

As seen in the above argument, for resonances with a wider width, it is more 
difficult to extract the resonance mass than the zero width state. The results
unavoidably have some ambiguities from the assumption of the shape of the spectral 
function and the width. Nevertheless, we can still estimate physical quantities
because of the following two points: 
(i) As illustrated in Figs.\ref{breit} and \ref{massrho},
although the width effects violate the stability of physical quantities,
the stability is still moderate than naively expected 
even if we take relatively large widths, 400 MeV. 
Thus, even if the width exists, the usual determination
of $E_{th}$ may work as a first approximation.
(ii)
As an experience from QSR for mesons and baryons with small widths, 
QSR employing $E_{th}$ taken around the position of the second resonances
does satisfy the least sensitibity criterion.
Since it is known experimentally that
the continuous cross section begins from around second resonance,
to fix $E_{th}$ around the second resonance is reasonable
and seems to be consistent with the philosophy of the quark-hadron
duality.
Therefore, we can expect the reasonable range of $E_{th}$
using the experimental facts.

%

%
%%%%%%%%%%%%%%%%%%%%%%%%%%%%%%%%%%%%%%%%%%%%%%%%%%%%%%%
\section{SU(3) singlet and octet states in the chiral SU(3) limit}
\label{sec:sinletoctet}
The $\sigma$ meson may be realized as admixture 
of the flavor singlet and octet states, and the $f_{0}(980)$ meson will be
its nonet partner of the flavor mixing in the flavor SU(3) symmetry breaking.
It is important for the studies of the $\sigma$ and $f_{0}(980)$ meson 
to know whether both resonances should be reproduced in the same footing
and what makes the differences.
For this purpose, we investigate quark flavor contents of 
the correlation functions for the flavor singlet and octet states 
in the SU(3) chiral limit, in which they decouple each other and 
we can investigate these states separately.
In the SU(3) chiral limit, 
we will find that the splitting of the singlet and octet
states stems only from the annihilation diagrams in which some quark
lines are not connected between the space-time points $x$ and $0$
in the intermediate states of the correlation function. There are 
two types of the annihilation diagrams in the tetraquark case, 
the color singlet two quark state, $(q\bar{q})_1$,
and the color octet quark with gluon,
$(q\bar{q})_8 G$. We will emphasize that 
the contributions from the $q\bar{q}$ annihilation diagrams are
more important in the tetraquark correlator than
in two quark meson correlator.

\subsection{Quark contents of the singlet and octet states}
To study the quark contents of
the light scalar nonets,
it is convenient to introduce the diquark basis,
\begin{eqnarray}
U=(\bar{d}\bar{s}),\ D=(\bar{s}\bar{u}),\ S=(\bar{u}\bar{d}),
\end{eqnarray}
which are anti-symmetrized in the color and flavor spaces.
We do not specify the Lorentz structure in this section,
since the flavor contents are the main issue here. The details of the
interpolating fields will be given in Sec.\ref{sec:IF}.
In the tetraquark case, 
the quark content of the singlet state is described
in the diquark basis as
\begin{equation}
{\cal S} = \frac{1}{ \sqrt{3} } ( U\bar{U} + D\bar{D} + S \bar{S}),
\label{eq:quarkcontentssinglet}
\end{equation}
and for the octet states, for example, the quark contents are given by
\begin{eqnarray}
{\cal O}_{1} &=& U \bar{D} \nonumber \\ 
{\cal O}_{2} &=& \frac{1}{ \sqrt{6} } ( U\bar{U} + D\bar{D} -2 S \bar{S}),
\label{eq:quarkcontentsoctet1}\\
 \dots && \nonumber
\end{eqnarray}
in a similar way to the quark contents
of the usual $q\bar{q}$ meson cases. 

The isodoublet $\kappa$ and isovector $a_0$ belong to purely
the octet because of the nonzero isospin,
while the isoscalar $\sigma$ and $f_0$ can be composed of the mixture
of the singlet and octet states 
in the real world 
where the flavor SU(3) symmetry is broken by the quark masses.
If the ideal mixing is realized,
the $\sigma$ and $f_{0}$ can be written as
\begin{eqnarray}
\sigma &\sim& (ud)(\bar{u} \bar{d}) = \bar{S} S 
 = \sqrt{ \frac{1}{3} } {\cal S}
- \sqrt{ \frac{2}{3} } {\cal O}_{2},
\label{sigma} \\
f_0 &\sim& \frac{1}{ \sqrt{2} }
 [ (us)(\bar{u}\bar{s}) + (ds)(\bar{d}\bar{s}) ] \nonumber \\
 &=& \frac{1}{ \sqrt{2} }[ \bar{D}D + \bar{U}U ]
  =  \sqrt{ \frac{2}{3} } {\cal S}
   + \sqrt{ \frac{1}{3} } {\cal O}_{2}. 
\label{f0}
\end{eqnarray}

From now on, 
we consider the chiral limit taking the zero quark masses.
This limit is convenient to study the difference between 
the singlet and octet states.
The study of the sigma meson in this work is valid even in the 
physical world with the finite strange quark mass, since, with the SU(3)
breaking due to the strange quark mass, the ideal mixing is expected 
to be realized and the quark contents of the sigma meson is given by
Eq.\eqref{sigma}. In this case, the sigma meson consists  only 
of the up and down quarks. 
For the $f_{0}$ and the octet mesons, it is important to include the finite 
strange quark mass.  

We will find that important contributions for the difference between
the singlet and octet states are
the mixing of four-quark (4q) states with 
two-quark (2q) states, two-quark and gluon mixed (2qG) states
or glueball states.

\subsection{The annihilation diagrams in 2q-2q and 4q-4q correlators}
As seen in Eqs.\eqref{sigma} and \eqref{f0}, the difference of the singlet
and octet correlation functions is important to make the splitting of the $\sigma$
and $f_{0}$ mesons. Here we would like to focus on the annihilation 
diagrams, 
since they are only the source of the splitting.
We will also see that the 
annihilation diagrams play key roles especially 
for the scalar meson spectra.
The remarkable point is that the effects of the annihilation diagrams are more
important in the 4-quark correlators than the 2-quark mesonic correlators.
This can be understood from the viewpoint of the momentum conservation.
Here the annihilation diagram means the diagrams 
which have some quark lines disconnected between $x$ and $0$.

\begin{figure}[floatfix]
  \begin{center}
   \resizebox{75mm}{!}{\includegraphics{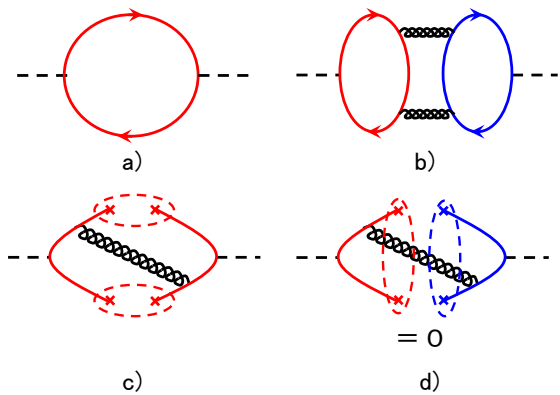}}
    \caption{(Color online) Some diagrams in the 2-q correlators.
   The flavor is distinguished by the line of color.
   a) The leading perturbative contribution in dim.0 term.
   b) The perturbative $q\bar{q}$ annihilation diagrams in dim.0 term
   with suppression factors, $1/N_c$ and $\alpha_s^2$.
   c) The leading power correction in dim.6
   including $(q\bar{q})_8 G$ configurations without annihilations.
   d) The O$(\alpha_s)$ power correction in dim.6 to the annihilation diagrams
      is vanished because of traceless properties of Gell-Mann matrix. }
    \label{2q}
   \vspace{-0.5cm}
  \end{center}
\end{figure}

First of all, we start with
discussion about the annihilation diagrams in the two-quark (2q-2q) correlators.
Some typical OPE diagrams are shown in Fig.\ref{2q}.
The dashed circles represent a pair of quarks to form the condensate.
The perturbative diagrams shown in Fig.\ref{2q} a) and b)
contribute relatively high energy side
of the correlation function.
Diagram a) includes 1-loop without $\alpha_s$ corrections.
Diagram b) is one of the annihilation diagrams, since the quark lines are
not connected in the intermediate state. 
This diagram
has three loops with $\alpha_s^2$ correction
and consequently it is strongly suppressed compared to diagram a).
In diagram b) the two-gluon propagation
is necessary because of the traceless property of the color SU(3) matrix.
The suppression of this type of the diagrams can be also explained
from large $N_c$ counting \cite{Witten}.
The higher dimension terms shown in Fig.\ref{2q} c) and d)
contribute to the low energy side of the correlation funciton.
Diagram c) is not the annihilation diagram since the quark lines are connected 
through the quark condensates. Diagram d) is very similar with diagram c),
but the pairs of the quarks to form the condensate are different. Diagram d) is
the annihilation diagram. However, this contribution vanishes
at this O$(\alpha_s)$ order because of the 
color traceless properties.
All of these facts suggest the strong suppression relative
to the non-annihilation diagrams
from low to high energy
in the case of the 2q-2q correlators.

\begin{figure}[floatfix]
  \begin{center}
   \resizebox{75mm}{!}{\includegraphics{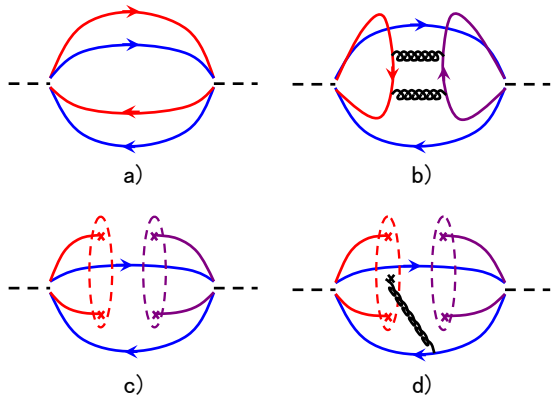}}
    \caption{(Color online) Some diagrams in the 4-q correlators.
   The flavor is distinguished by the line of color.
   a) The leading perturbative contribution in dim.0 term.
   b) The leading $q\bar{q}$ annihilation diagram in dim.0 term.
   c) The leading power correction in dim.6 to the annihilation
   diagrams including $(q\bar{q})_1$ configuration without
   $\alpha_s$ suppression factor. 
   d) The annihilation diagrams in dim.8
   which includes the configuration $(q\bar{q})_8G$.}
   \label{4q}
   \vspace{-0.5cm}
  \end{center}
\end{figure}

Next we turn to the discussion of the annihilation diagrams
in the four-quark (4q-4q) correlators.
We show in Fig.\ref{4q}
some examples of the diagrams in the 4q-4q correlators.
Diagrams a) and b) are perturbative diagrams. 
Diagram b) is the annihilation diagram but it is suppressed by $\alpha^{2}_{s}$
in the same way as the 2q-2q correlator case.
Diagram c) is the leading annihilation diagram with $(q\bar{q})_1$ configuration.
In contrast to the 2q-2q correlator cases, the $q\bar{q}$ annihilation diagrams
can appear without gluon propagations,
since the other two quarks can carry the momentum. 
Consequently no $\alpha_{s}$ suppression comes in. 
Diagram d) is also the annihilation diagram 
in dim.8 with the configuration $(q\bar{q})_8G$.
This is not again suppressed by the $\alpha_{s}$ correction.
Such annihilation diagrams can widely appear
in the higher dimension terms. This means that 
the annihilation processes
are much more important in lower energies where
we would like to investigate the properties of the scalar mesons. 

\begin{figure}[floatfix]
  \begin{center}
   \resizebox{60mm}{!}{\includegraphics{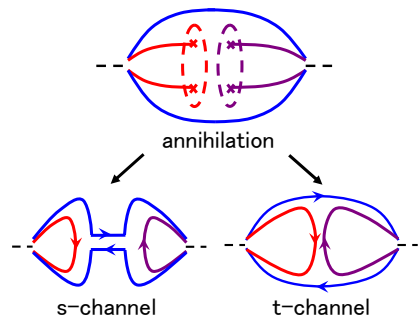}}
   \caption{(Color online) The two ways of interpretation
   of the annihilation diagrams.
   By deforming the quark line,
   we can interpret the annihilation diagrams as 
   either 2q like s-channel propagation 
   or 4q like t-channel propagation with
   exchange of mesonic resonances.
   }
   \label{anni}
   \vspace{-0.5cm}
  \end{center}
\end{figure}

Here we would like to stress that
the annihilation processes do not directly
imply 2q propagation emerged from tetraquark configuration.
Indeed, as shown in Fig.\ref{anni},
the annihilation diagrams can represent
either 2q s-channel propagation or 4q propagation
with exchanging the resonance in the t-channel.
Especially, contributions from the latter processes
can be interpreted
as diquark-diquark correlation
in our tetraquark fields as product of diquark fields.
We will see later that the annihilation diagrams 
provide imporant contributions in our Borel analyses.

\subsection{The splitting between singlet and octet states}
\label{sec:splitting}
Now we discuss how the annihilation diagrams studied in the previous 
subsection contribute to the flavor singlet and octet correlation functions.
The flavor contents of the singlet and octet correlators are given
respectively by
\begin{eqnarray}
&&\!\!\!\!\!\!\!\!\!\!\!\!\!\!\!\!
\frac{1}{3} \big\langle T \big[ \bar{U}U(x) \big]
 \big[ \bar{U}U(0) 
 + \bar{D} D(0)
 + \bar{S} S(0) 
\big] \big\rangle  + ({\rm perm.}), \label{eq:singlet}\\
&&\hspace{2cm} 
\big\langle T \big[ \bar{U}D(x) \big] \big[ \bar{U}D(0) \big]
\big\rangle, ... \label{eq:octet}
\end{eqnarray}
where for the octet we use Eq.\eqref{eq:quarkcontentsoctet1}.
These correlators consist of 
three types of correlators in terms of the flavor content:
\begin{eqnarray}
&& \big[ \bar{U}U(x) \big] \big[ \bar{U}U(0) \big]
  =  \big[ ds\bar{d}\bar{s}(x) \big] \big[ \bar{d}\bar{s}ds(0) \big],
  \label{uuuu}\\
&& \big[ \bar{U}U(x) \big] \big[ \bar{D}D(0) \big] 
 = \big[ ds\bar{d}\bar{s}(x) \big] \big[ \bar{s}\bar{u}su(0) \big],
 \label{uudd}\\
&& \big[ \bar{U}D(x) \big] \big[ \bar{U}D(0) \big]
 = \big[ ds\bar{s}\bar{u}(x) \big] \big[ \bar{d}\bar{s}su(0) \big]
 \label{udud}.
\end{eqnarray}

In the first expression, all the diquarks have the same flavor,
the second expression is the correlation between the different flavor 
tetraquarks consisting a pair of the same flavor diquarks, and the third
expression has two same tetraquarks which are made up of a pair of
different diquarks.
For the flavor singlet, the first two contribute, while for the octet only
the last type of the correlators takes place.

Now we count possible configurations for the annihilation diagrams 
in each correlator shown above. Since neither QCD interaction nor
vacuum condensates can change quark flavors, we can connect only
the pair of the quark and antiquarks which have the same flavor. Thus, 
the first correlator \eqref{uuuu} has both the four-quark connected 
diagrams like diagram a) in Fig.\ref{4q} and the $q \bar q$ 
annihilation diagrams like diagram c) and d) in Fig.\ref{4q}:
\begin{eqnarray}
&& ds\bar{d}\bar{s} \rightarrow ds\bar{d}\bar{s}\ \  ({\rm direct}),
\\
&& ds\bar{d}\bar{s} \rightarrow 
[ (d\bar{d})_1,(d\bar{d})_8 G, G^2 ]\rightarrow ds\bar{d}\bar{s},
\label{ddG}\\ 
&& ds\bar{d}\bar{s} \rightarrow
[ (s\bar{s})_1,(s\bar{s})_8 G, G^2 ] \rightarrow ds\bar{d}\bar{s},
\label{ssG}\ \ 
\end{eqnarray}
where we explicitly write down the possible color singlet
configurations for the annihilation processes in the intermediate states.
Here we have also mentioned the  $q\bar{q}q\bar{q}$ annihilation 
process as $G^{2}$ in the intermediate state. This process 
gives 4q-glueball correlation. 
In leading $\alpha_s$ analysis up to dim.12, however,
such $qq\bar{q}\bar{q}\rightarrow G^2$ diagrams do not appear.
In the similar way, we find that 
the second type of the correlator \eqref{uudd} has only the 
annihilation diagrams:
\begin{eqnarray}
ds\bar{d}\bar{s} \rightarrow 
[(s\bar{s})_1, (s\bar{s})_8 G, G^2] \rightarrow su\bar{s}\bar{u}.
\label{dsG}
\end{eqnarray}
The last correlator, which contributes to the octet, has both
the four quark connected diagrams and the $q\bar q$ annihilation
diagrams:
\begin{eqnarray}
&&ds\bar{s}\bar{u} \rightarrow ds\bar{s}\bar{u}\ \  ({\rm direct}),\\ 
&&ds\bar{s}\bar{u} \rightarrow 
[(d\bar{u})_1, (d\bar{u})_8G] \rightarrow ds\bar{s}\bar{u},
\end{eqnarray}

As we have mentioned, the first and second types of the correlators
contribute to the singlet state, while the octet correlator has only the third one. 
From these observations, now we can see the singlet state
has much more annihilation diagrams, or 4q-2q
correlation effects than the octet states.
In fact, the explicit calculation given in the Appendix.\ref{sec:appA}
shows the number of the $q\bar{q}$ annihilation diagrams
is larger in singlet than in octet by factor 4.

\section{The Borel analyses}
\label{SecBorel}
In this section,
we perform the Borel analyses for the $\sigma$ meson and also 
the scalar states with the flavor singlet and octet in the flavor SU(3) limit.
We first discuss the interpolating fields used in these analyses in
subsection \ref{sec:IF}. 
There we introduce linear combinations of the 
interpolating fields constructed by the scalar and pseudoscalar diquarks
with a mixing angle $\theta$. We also discuss the criterion to determine
the interpolating 
fields based on the argument for the Borel window given in subsection \ref{sec:Basics}.
In subsection \ref{sec:SingletOctet}, we show the results for the flavor singlet and
octet states. 
Subsection \ref{sec:sigmaSR} is devoted to discussion for the result of 
the $\sigma$ meson. 
There we will find that the correlation function 
in the Borel analysis shows large contributions around $E=600\sim800\ {\rm MeV}$.

Throughout our analyses,
we use the standard values of the condensates in OPE: $\langle \bar{q}q \rangle=-(0.230\ {\rm GeV})^3$,
$\langle \bar{q} g_s \sigma G q \rangle/ \langle \bar{q}q \rangle = 0.8\
{\rm GeV}^2$,
$\langle \frac{\alpha_s}{\pi} G^2 \rangle = 0.012\ {\rm GeV}^4$
and $\alpha_s(1{\rm GeV})=0.4$.
The OPE is evaluated within the factorization hypothesis
without quantum loop corrections.
The detail of the OPE calculation is given in Appendix.\ref{sec:appA}.

\subsection{Interpolating fields}
\label{sec:IF}
In QSR, it is technically important to use such good interpolating fields
as to pick up largely  the resonance state
and have smaller correlation with the higher energy states.
With the number of quarks increasing, there are several choices of
the local interpolating fields with the quantum number of our interest 
in contrast to the $q \bar q$ meson cases.
The authors of Ref.\cite{rcnp}
comprehensively studied the possible local interpolating fields
of the tetraquark and the properties of the scalar mesons by
constructing QCD sum rules with OPE up to dim.8.  
In this work, 
we calculate the OPE of the correlation function up to dim.12, 
considering linear combinations of the following two operators for the $\sigma$ meson:
\begin{eqnarray}
J_P &=& \epsilon^{abc} \epsilon^{dec} 
 [u_a^T C d_b][\bar{u}_d C \bar{d}_e^T], \\
J_S &=& \epsilon^{abc} \epsilon^{dec}
 [u_a^T C\gamma_5 d_b][\bar{u}_d \gamma_5 C \bar{d}_e^T], 
\end{eqnarray}
where $a,b,c...$ represent the color indices, 
$\psi^{T}$ means the transposed spinor of $\psi$ and $C$ is the charge
conjugation matrix. 
The above interpolating fields, $J_P$ and $J_S$, are constructed by two pseudoscalar diquarks and
two scalar diquarks, respectively.
The  linear combination is given with a mixing angle $\theta$ $(0\le \theta < \pi)$ by
\begin{eqnarray}
J(\theta) = \cos\theta \, J_P + \sin \theta\,  J_S.  \label{eq:intapo}
\end{eqnarray}

The mixing angle $\theta$ in the interpolating field \eqref{eq:intapo} 
is determined so that the interpolating field couples more strongly to
the resonance state 
and have less contributions from the higher energy states
and scattering background, 
which is achieved by imposing the following criteria:
\begin{enumerate}
\renewcommand{\theenumi}{\bf\alph{enumi})}
\item Sufficiently wide Borel windows are established in the sum rule.
\item The effective masses and residua are weakly dependent on the Borel mass $M^{2}$ as much as possible.
\item The results have also adequately weak dependence on the threshold $E_{th}$.
\item The effective residua are satisfactorily large.
\end{enumerate}
These criteria are not independent each other and  
the mixing angle can be insensitive to some of them. 
Hence, the mixing angle is not uniquely fixed in 
quantitative manner.
Here we discuss typical mixing angles which simultaneously satisfy the above
criteria with better degree.
In the followings, we explain the meanings of the criteria separately
and understand the priorities among these criteria.

The criterion {\bf a)} is most important, being essential to avoid the pseudo peak artifacts
and reduce truncation errors of the OPE, as emphasized in Sec.\ref{sec:Basics}.
Thus, this criterion should be always satisfied independently of 
the other 
criteria to guarantee the sum rules contain largely contributions from low energy physics. 

After we find the Borel window in the sum rules,
the next task is to establish better pole isolation
from the background contamination,
which can be done by satisfying the criteria {\bf b)} and {\bf c)}.
The background contamination comes mainly from 
scattering states of two mesons and other resonance states. 
In the low energy side,
scattering states of two Nambu-Goldstone bosons,
i.e., $\pi\pi$, $KK$, $\eta \pi$, scattering states
can appear from $E=0$ in the present analyses because 
all the Nambu-Goldstone bosons are massless in the chiral limit.
If these states largely contribute to the sum rules,
the effective residue and mass obtained in the sum rules 
have larger $M^2$ dependence in the $M^{2}$ region below the resonance pole
than we expect from the resonance width, which has been
discussed in Sec.\ref{sec:width}.
This contamination below the resonance pole can be reduced by the selection of 
$\theta$ with satisfying the criterion {\bf b)} well.
The background contamination above the pole energies comes 
from higher excited resonance states and meson scattering states 
other than the Nambu-Goldstone bosons.
In the case that such contamination is large, 
the physical quantities have large dependence against the variation of $E_{th}$.
This can be reduced by imposing the criterion~{\bf c)}.

The criterion {\bf d)} is imposed to have the large overlap of 
the interpolating field with the resonance pole and to reduce the
truncation errors of OPE. 
In our analyses, 
we neglect higher-dimensional terms of OPE than the dim.12 terms.
If the residua obtained in the sum rule are sufficiently large, the interpolating fields
used in the analyses have enough overlaps to the resonance states and the sum
rules constructed with the approximated OPE may contain so much information for the resonance to be extracted. But,
if the obtained residua are small, the neglected higher-dimensional terms might have
important information on the resonance states, although the OPE convergence
is well provided. Thus, we consider the criterion {\bf d)}. 
Nevertheless, the criterion {\bf d)} has less priority than the criteria {\bf b)} and {\bf c)}.
This is because the best choice of $\theta$ is not always
the one which have the largest overlap with the region below $E_{th}$.
Even if the low energy correlation is large,
it can be just a signal of strong contamination from the background.
Therefore, the most important thing is that we construct the sum rules
within the sufficiently wide Borel window satisfying the criterion {\bf a)}
with the best choice of the interpolating field which 
have the least overlap with the background.

%%%%%%%%%%
\subsection{Borel analyses of flavor singlet and octet states}
\label{sec:SingletOctet}
%%%%%%%%%%

Let us consider the flavor singlet and octet states produced by the interpolating 
fields given in Eq.\eqref{eq:intapo} with the mixing angle $\theta$ and
having the quark contents \eqref{eq:quarkcontentssinglet} and
\eqref{eq:quarkcontentsoctet1}.
The mixing angle is determined by the criteria discussed in Sec.\ref{sec:IF}. The 
detailed discussions and the technical issues are given in Appendix.\ref{sec:thetadep}. 
Our findings for the mixing angle are as follows:
\begin{itemize}
  \item the Borel windows can be established in all the mixing angles except $\theta \sim 0$ for the flavor octet case. 
  \item the results for the physical quantities are not very sensitive to the choice of the mixing 
  angle $\theta$ except some region of $\theta$ both in the flavor singlet and octet cases.
  \item one of the best value of the mixing angle both for the singlet
	and octet is $\theta \simeq 7\pi/8$.
	Thus in this section we will show the results for $\theta=7\pi/8$.
  \item the residua of the singlet states have larger values than those
	of the octet states in almost all region of $\theta$.
  \item the effective masses of the octet case are typically smaller
	than those of the singlet, and the Borel stability is worse in the octet case. 
\end{itemize}
The last two findings may imply, in comparison to the singlet case, that
the octet interpolating field used here 
does not have sufficiently large overlap
with the resonance state or that the width of the resonance in the octet
channel is considerably large. For further investigation of the octet states,
it might be necessary to use interpolating fields with different Lorenz structure. 

\begin{figure}[floatfix]
  \begin{center}
   \resizebox{87mm}{!}{\includegraphics{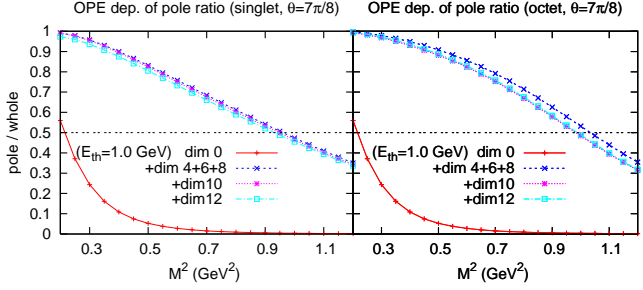}}
   \caption{(Color online) The OPE truncation dependence of the pole dominance, $B(M^2;s_{th})$,
   for singlet and octet case with the threshold $E_{th}=1.0$ GeV. }
   \label{sodimconpole}
  \end{center}
\end{figure}

\begin{figure}[floatfix]
 \begin{center}
   \resizebox{87mm}{!}{\includegraphics{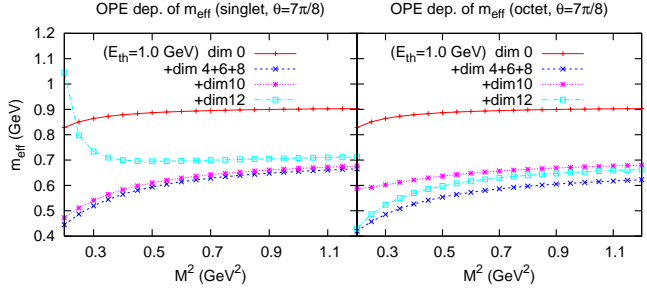}}
  \caption{(Color online) The OPE truncation dependence of the effective masses for
  singlet and octet case with the threshold $E_{th}=1.0$ GeV.}
  \label{sodimmass}
 \end{center}
\end{figure}

Hereafter we proceed to
discussions of the Borel analyses of the scalar mesons
with fixing the mixing angle as $\theta = 7\pi/8$. 
First of all, we examine 
the OPE convergence of the pole dominance, 
which are important to establish the appropriate Borel window, 
and we emphasize importance of the higher dimension terms. 
Here we fix the threshold to a typical value $E_{th}=1.0$ GeV.
We show in Fig.\ref{sodimconpole}  the OPE dependence of
the ratio of the pole contribution to the whole spectral function
$B(M^2;s_{th})$ defined in Eq.(\ref{poleratio})
for singlet and octet cases.
We can see that the OPE corrections to the perturbative term (dim.0)
is extremely important to achieve the pole dominance. 
It can be also seen in Fig.\ref{sodimconpole} that 
the higher dimension terms than dim.8 
is not so important for the pole dominance.
Nevertheless, as seen in the plots of the effective masses (Fig.\ref{sodimmass}),
the dim.10 and 12 terms are important to obtain the Borel stability 
in the small $M^2$ region
(typically $M^2<1.0\ {\rm GeV^2} \sim M_{ {\rm max} }^2$)
where we will evaluate the physical quantities later.
This means that these higher dimension terms have essential to reproduce
the low energy resonances. 
For the singlet case, the dim.12 
term improves the Borel 
stability well, while in the effective mass for the octet state the
dim.10 and 12 terms do not help as well as the singlet
case. The Borel stability obtained with including only the dim.0 term
can be interpreted as the artificial stability discussed in Sec.\ref{sec:artifact},
since the pole dominance is not achieved in these cases. 
As emphasized in Sec.\ref{sec:splitting}, the difference of 
the effective masses of the singlet and octet comes
from  the annihilation diagrams, which is increasingly important
in higher dimension terms or low energy region.


\begin{figure}[floatfix]
  \begin{center}
   \resizebox{87mm}{!}{\includegraphics{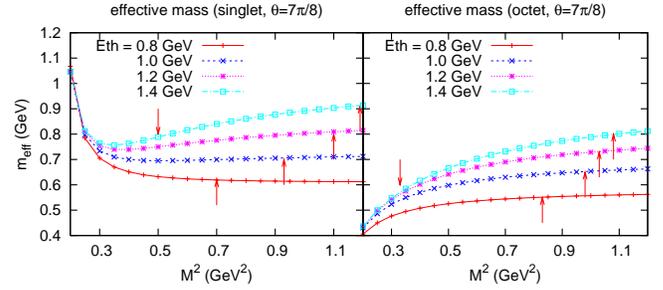}}
   \caption{(Color online) 
   The effective masses for the singlet and octet states.
   The values of the threshold $E_{th}$ are $0.8,\ 1.0,\ 1.2$ and 1.4 GeV.
   The downward and upward arrows indicate
   the lower and upper bounds of the Borel window,
   respectively. }
   \label{somass}
  \end{center}
\end{figure}

Next we evaluate the resonance masses and residua 
for the flavor singlet and octet states in the chiral limit. 
We calculate the masses and residua widely with
several values of $E_{th}=0.8$, $1.0$, $1.2$, and $1.4$ GeV,
in order to see the threshold dependence discussed in Sec.\ref{sec:IF} and
the effect of the resonance width studied in Sec.\ref{sec:width}. 
We show in Fig.\ref{somass} the plots of 
the effective mass for singlet and octet states
with $E_{th}=0.8$, $1.0$, $1.2$, and $1.4$~GeV.
The downward and upward arrows indicate
the lower and upper bounds of the Borel window,
respectively. The way to fix the Borel window has been
given in Eqs.\eqref{opeconv}  and \eqref{poleratio}. 
The lower Borel mass is not dependent on the threshold,
since it is determined only by the OPE convergence. 

We extract the masses of the scalar mesons from the effective
mass plot in Fig.\ref{somass} with consideration of its possible modifications
from the resonance width, as discussed in Sec.\ref{sec:width}.
For the singlet state, we find  the moderate Borel stability
for $E_{th}=1.0\sim1.4$ GeV in Fig.\ref{somass}.
This would imply that some resonance state saturate the spectral 
function around $E\sim 700$ MeV. 
We estimate the mass for the singlet state at 700 MeV for a small width
state, finding almost perfect Borel stability with $E_{th}=1.0$ GeV.
If the state has some larger width, we evaluate the mass by 850 MeV 
with a possible width 400 MeV according to the discussion in Sec.\ref{sec:width},
in which we found that the Borel stability is not perfectly achieved in the case
of the resonance with width. Finally we conclude that the present 
QCD sum rule estimate the flavor singlet scalar meson by $700\sim 850$~MeV
with consideration of the possible width up to 400~MeV. 

For octet state, we estimate the resonance mass at $600\sim 750$~MeV in the 
similar way to the singlet case. Nevertheless, the effective mass in the octet channel depends on the Borel mass fairly.
One of the possible explanations is that
the signal of the octet resonance is weak and
considerably affected by low energy scattering states.
Another possibility is that the observed state has the large width, let us say,
$100\sim400$ MeV, which can be estimated 
from the comparison to the effective mass of 
the Breit-Wigner form in Fig.\ref{breit}.
Probablly both effects may play important roles in the octet sum rules.

\begin{figure}[floatfix]
  \begin{center}
   \resizebox{87mm}{!}{\includegraphics{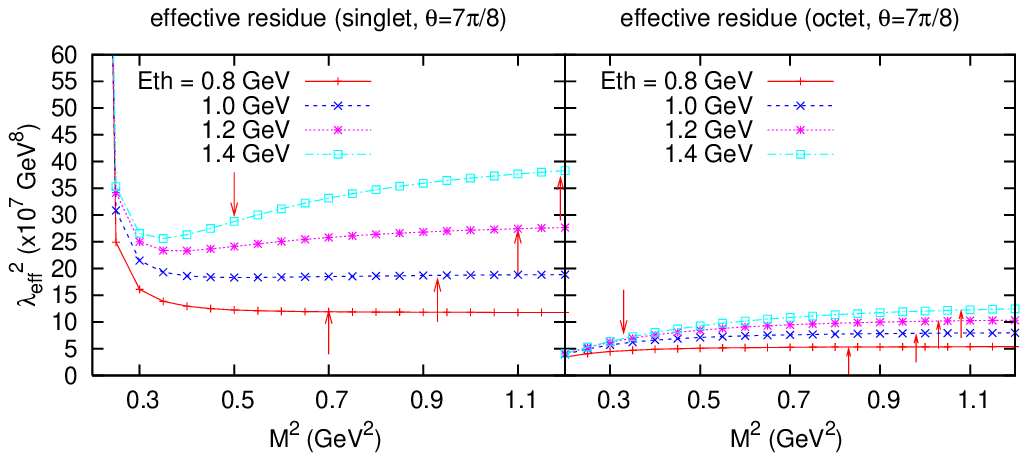}}
   \caption{(Color online)
   The effective residua for the singlet and octet states.
   The values of the threshold $E_{th}$ are $0.8,\ 1.0,\ 1.2$ and 1.4 GeV.
   The downward and upward arrows indicate
   the lower and upper bounds of the Borel window,
   respectively. }
   \label{soresi}
  \end{center}
\end{figure}

We also plot the effective residua for the singlet and octet cases in Fig.\ref{soresi}.
The Borel stability is achieved as similarly as that of the effective mass.
The residua for single and octet states
are evaluated by $(20\sim 35)\times 10^{-7}\ {\rm GeV^8}$ and 
$(7.5\sim 13)\times 10^{-7}\ {\rm GeV^8}$, respectively. 
The overlap strength of the singlet case is factor $2\sim3$ larger than
the octet state.

Both in the singlet and octet states, the threshold dependence is not very small.
This would suggest the separation between resonance pole
and threshold would not be completely done. For the further investigation, 
we would need more elaborated technique to isolate the resonance states,
for instance, large $N_{c}$ argument as discussed in Ref.\cite{KHJ2}.

\subsection{Borel analysis for sigma meson}
\label{sec:sigmaSR}
We discuss the Borel analysis for the $\sigma$ meson in the chiral limit. 
With the SU(3) breaking, 
the ideal mixing of the flavor singlet and octet components may
be realized and the lighter scalar meson with $I=0$ and $S=0$, 
namely the $\sigma$ meson, 
may be described by only the up and down quarks. 
(The ideal mixing is the assumption in this work. 
To check this assumption, we need to investigate the flavor mixing angle
in QCD sum rules with a finite strange quark mass.) 
For the ideal mixing, the spectral function for the $\sigma$ meson is given 
by a linear combination of the singlet and octet spectral functions in the chiral limit: 
\begin{eqnarray}
\langle T J_{\sigma}\bar{J}_{\sigma} \rangle 
= \frac{1}{3}\langle T J_{ {\cal S} }\bar{J}_{ {\cal S} } \rangle
+ \frac{2}{3}\langle  T J_{ {\cal O} }\bar{J}_{ {\cal O} } \rangle.
\label{superposition}
\end{eqnarray}
We work again with $\theta = 7\pi/8$ for the mixing angle of the interpolating field.

\begin{figure}[floatfix]
  \begin{center}
   \resizebox{85mm}{!}{\includegraphics{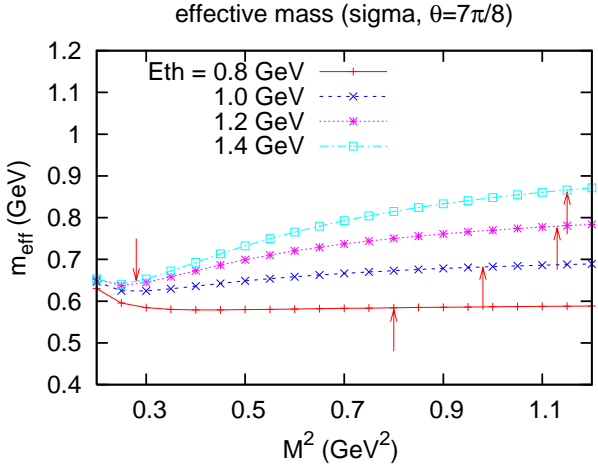}}
   \caption{(Color online) The effective mass for the $\sigma$ meson.
   The values of the threshold $E_{th}$ are $0.8,\ 1.0,\ 1.2$ and 1.4 GeV.
   The downward and upward arrows indicate
   the lower and upper bounds of the Borel window,
   respectively. }
   \label{sigmamass}
  \end{center}
\end{figure}

\begin{figure}[floatfix]
  \begin{center}
   \resizebox{85mm}{!}{\includegraphics{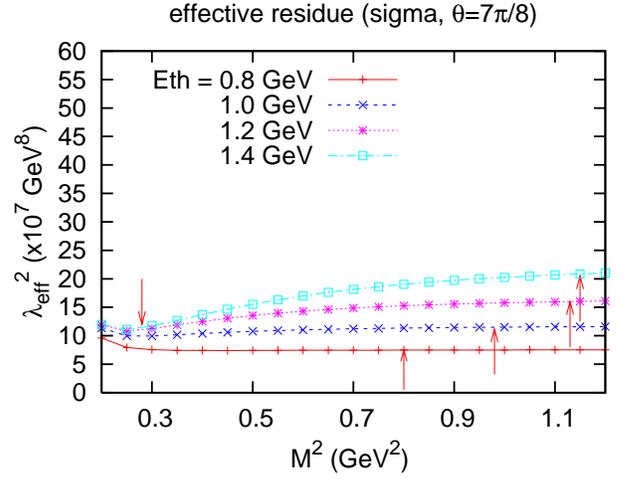}}
   \caption{(Color online) The effective residue  for the $\sigma$ meson.
   The values of the threshold $E_{th}$ are $0.8,\ 1.0,\ 1.2$ and 1.4 GeV.
   The downward and upward arrows indicate
   the lower and upper bounds of the Borel window,
   respectively. }
   \label{sigmaresi}
  \end{center}
\end{figure}

Evaluating the effective mass and residue of the sigma meson, we obtain the plots with
$E_{th}=0.8$, 1.0, 1.2 and 1.4~GeV in Figs.\ref{sigmamass} and \ref{sigmaresi}.
In these plots, we find again fairly good Borel stability both for the effective mass
and residue. 
In the same way as the flavor singlet case, 
we estimate the resonance pole mass at $600\sim 800$ MeV 
with paying attention to the possible width $\sim 400$ MeV.
The pole residue is also evaluated by $(10\sim 20)\times 10^{-7}\ {\rm GeV^8}$.
The value of the residue is consistent with the expectation that it is a superposition 
of the singlet and octet states with group theoretical weights for the ideal mixing,
which are 1/3 and 2/3 for the singlet and octet contributions, respectively. 

This $\sigma$ mass obtained in the present sum rule analysis with 
the tetraquark operator is considerably smaller than that extracted 
from correlator analyses with the two-quark operators, 
which is $\sim 1.2\ {\rm GeV}$.
As discussed in Sec.\ref{sec:SingletOctet},
the 2q operator mainly couples to the 2q components,
while the tetraquark operators induce both the 4q components 
and the considerable 2q and 2qG components. 
The present tetraquark investigation shows that 
that inclusion of 4q and 2qG components is relevant to explain
the smaller  $\sigma$ mass~\cite{TITech}. 

As a whole, finding the wider Borel window in the sigma meson sum rule, 
we conclude that the results for the $\sigma$ meson is more reliable than the singlet and octet
cases. 
Especially it is notable that the lower bound of the Borel mass is sufficiently small.
This helps us to investigate low energy contributions for the $\sigma$ channel. 
Due to the small lower bound of the Borel mass and the Borel stability obtained 
even in $M^2=0.3$ MeV$^2$, we conclude that the pole dominance of the $\sigma$ resonance
is fairly established with small contamination of low energy scattering. 
We also see that the large strength comes mainly from 
the flavor singlet component, which shows the large enhancement
around $E=700\sim 850\ {\rm MeV}$ region. 

In closing this section, we would like to emphasize that, although having
some ambiguities coming from the influence of the scalar meson width
on the Borel stability and the choice of the mixing angle $\theta$, the present
QSR analyses with the tetraquark operators provide the scalar meson masses
less than 1 GeV, which cannot be produced in QSR with 
2q operators.

%%%%%%%%%%%%%%%%%%%%%%%%%%%%%%%%%%%%%%%%%%%%%%%%%%%%%%%
\section{Discussion}

The present QSRs are constructed in the SU(3) chiral limit.
It is very interesting 
to extend our QSRs beyond the chiral limit with a realistic strange quark mass.
Especially for realistic calculation of the masses of the octet scalar mesons, 
such as $\kappa$ and $a_{0}$, inclusion 
of the strange quark mass is absolutely necessary.
It is also interesting to 
investigate the mixing angle of the flavor singlet and octet interpolating 
fields with finite strange quark mass. 
In this work, we have assumed the
ideal mixing for the flavor content of the $\sigma$ meson operator. 
But it is not trivial to realize the ideal mixing in the physical scalar meson nonet. 
Analyses with the strange quark mass for the flavor mixing angle is possible
within the QSR approach. 
This would clarify the strange contents of the $\sigma$ meson. 

Inclusion of the quark masses with the SU(3) symmetric way is also 
a good theoretical issue. 
In the present analyses we have found that
contributions from the annihilation diagrams are responsible for the
splitting of the singlet and octet states. 
It is good to know the
quark mass effects on the splitting. 
Especially, in the chiral limit,
we have found that the octet tetraquark operator weakly couples 
to the physical state, so that it is interesting to investigate how
the octet state changes with the finite quark mass. 

The issue of the flavor mixing angle and splitting of the singlet
and octet states is closely related to the mass difference of the $f_{0}$
and $a_{0}$. 
In the present experiments they almost degenerate. 
The $a_{0}$ meson may be classified into the flavor octet, while the 
$f_{0}$ meson may be described by a linear combination of the
singlet and octet operators with the flavor SU(3) breaking. 
Thus the mass difference is given by the mixing angle and contributions
of the annihilation diagrams in the QSR approach. 
It is interesting to
study how the degeneracy of $f_{0}$ and $a_{0}$ is realized in QSR. 

In this work, we have used the tetraquark operators with expectation 
to have strong coupling to the four-quark states and some contributions
from the two-quark (2q) and two-quark-one-gluon (2qG) states.
It is also interesting to investigate the scalar meson states using 
the operators having large overlap with two-quark states and 
small coupling with four-quark states and to compare this
with the present analyses. 
It has been reported in Ref.\cite{Narison}
that $\kappa$ and $a_0$ may have large components of
two-quark states rather than four-quark states and 
the coupling of the two-quark states with nonresonant scattering
states makes the $\kappa$ and $a_{0}$ masses lighter in a
similar way to the Feshbach resonances. 
In fact, the previous QSR studies with two-quark mesonic currents
for $\kappa$(800) and $a_0$(980) 
showed $1.0\sim1.2$ GeV and this is not too higher than
the experimental value. It would be natural to develope the ideas that
the 2q state with some additional components 
might lead the experimental mass.
In future,
we will report the analysis using such operator
with the help of the large $N_c$ argument \cite{KHJ2}.

In order to investigate further the widths of the scalar meson states, 
it is necessary to investigate three point correlation functions for the
$\sigma \rightarrow \pi\pi$, for instance. 
To obtain the decay width in QSR, 
we need to calculate both the two-point and three-point
correlation functions, and to perform combined analyses of them
in self-consistent ways to determine the mass, the residue, the  
threshold and the mixing angle. 

It is also interesting to calculate two photon decays of the scalar mesons
in QSR in order to compare mesonic molecule pictures. 
It has been reported
in Ref.\cite{Barnes} that the $K\bar K$ molecule picture of the $f_{0}$ and 
$a_{0}$ mesons proposed by Ref.\cite{Weinstein} provides a factor 3 larger
decay rate of $a_{0} \rightarrow \gamma\gamma$ than experiments 
and also that two quark picture
predicts much larger decay rates of $a_{0} \rightarrow \gamma\gamma$ and
$f_{0} \rightarrow \gamma\gamma$. 
The discrepancy
could be explained by the tetraquark picture, since decay of four quarks 
into two photons may be suppressed compared to two quarks to two photons
by the electromagnetic constant $\alpha$ and
the quark wavefunction $|\Psi(0)|^{4}$ at the origin. 

It is an important question whether correlation function analyses
with local operators, such as the present QSR, can reproduce
nature of spatially extended objects like mesonic molecular 
states.  The correlation functions observe overlaps
of the local operators and the physical states. In principle,
if the overlap is not extremely small, the correlation functions
have components of the molecular states. However, it is 
not sufficient at all to investigate the nature of the physical states only from
the overlap of the local operators and the physical states. 
We certainly need other information of the wavefunctions, such as
the decay widths.  

\section{Summary}
In this paper, 
we discuss the properties of the $\sigma$(600) meson
in the QCD sum rules (QSR) for the tetraquark operators,
emphasizing the importance of setting up the Borel window
to avoid artificial Borel stability.  
The flavor structure of the sigma interpolating field
is given by the SU(3) singlet and octet representations assuming
the ideal mixing with the SU(3) flavor breaking.
We invesigate also the flavor SU(3) singlet and octet states
for the light scalar meson with $I=0$ and $S=0$
in the massless limit with special attentions for the
roles of the constituents of the $\sigma$ meson.
Having shown that the splitting between the singlet and octet states
stems from the annihilation diagrams in which only two quark lines 
out of four carry the hard momenta, 
we have found that the contributions
from the annihilation diagrams are responsible for the mass
of the $\sigma$ state around $600\sim800$ MeV, which 
is much lighter than the mass extracted by QSR with 
the two quark operators. 

Investigation of the scalar meson in the present QSR have been 
carried out using the correlation functions for a linear combination
of the tetraquark operators made up by the scalar and pseudoscalar
diquarks. 
The mixing angle has been determined with QSR so as 
to achieve better resonance isolation from the background 
contamination.
We have also discussed
possible influence of 
the large width of the scalar meson on the Borel stability. 

For the singlet state, having found fairly good Borel stability within 
the practical Borel window, we have obtained the mass as
$700\sim 850$ MeV with considerations for the ambiguities coming from
the possible width of the scalar meson and small dependence on
the threshold parameter.
For the octet state, we have found rather poor Borel stability
and obtained a bit lighter mass around $600\sim 750$ MeV with smaller
coupling strength than the singlet case. 
This would suggest that, in the octet channel,
the signal of the resonance state is rather small compared to 
background scattering states of two Nambu-Goldstone bosons,
or that the resonance state has the large width.
The present analyses have been done in the SU(3) chiral limit. 
It is interesting to investigate the octet state beyond the
chiral limit with a realistic strange quark mass for the
the $\kappa$ and $a_0$ mesons.

For the $\sigma$ meson, having found wider Borel window 
and more Borel stability than the flavor singlet and octet
cases, we have estimated the resonance mass at $600\sim 800$ MeV
and the residue at $(10\sim 20)\times 10^{-7}\ {\rm GeV^8}$, which 
include again the ambiguities of the width and the choice of the threshold.
Although the present analyses have some ambiguities coming
from the influence of the physical width of the scalar meson on the
Borel stability and the choices of the threshold and the mixing angle, 
we conclude that the QSR with the tetraquark operators provides
the resonance masses of $600\sim 800$ MeV in the $\sigma$ meson
channel in the chiral limit.
This suggests that
the $\sigma$ meson has largely the four-quark components,
since the QSR with the 2q operators cannot produce masses
less than 1 GeV in the scalar channel.

\begin{acknowledgments}
We thank Professor H.\ Suganuma for helpful discussions.
This work is supported by the Grant-in-Aid
for the 21st Century COE ``Center for Diversity and Universality
in Physics'' from the MEXT of Japan and in part by 
the Grant for Scientific Research (No.\ 18042001).
This research is  part of Yukawa International Program for Quark-Hadron Sciences. 

\end{acknowledgments}

\appendix

\section{The OPE results}
\label{sec:appA}
In this appendix,
we present the expression of each term of the OPE
for the correlation function \eqref{eq:corrfunc} with the 
interpolating field \eqref{eq:intapo} having the mixing 
angle $\theta$ $(0\le \theta < \pi)$:
\begin{eqnarray}
J(\theta) = \cos\theta \, J_P + \sin \theta\,  J_S  
\end{eqnarray}
The interpolating fields with specific mixing angles have definite chiral
representations: 
$J(3\pi/4)$ is classified into the irreducible representation  
$[({\bf 3},\bar{\bf 3}) \otimes (\bar{\bf 3}, 
{\bf 3})]$ of the chiral SU(3)$_{L}\otimes$SU(3)$_{R}$ group, 
while $J(\pi/4)$ is a combination of the 
$[({\bf 8},{\bf 1}) \otimes ({\bf 1}, {\bf 8})]$ 
and $({\bf 1},\bar{\bf 1})$ representations. 
(Actually $J(\pi/4)$ is classified into the nonet 
$[({\bf 9},{\bf 1}) \otimes ({\bf 1}, {\bf 9})]$ representation of the
U(3)$_{L}\otimes$U(3)$_{R}$ group.) 

\begin{figure}[floatfix]
  \begin{center}
    \begin{tabular}{c}
      \vspace{0.0cm}
     \resizebox{85mm}{!}{\includegraphics{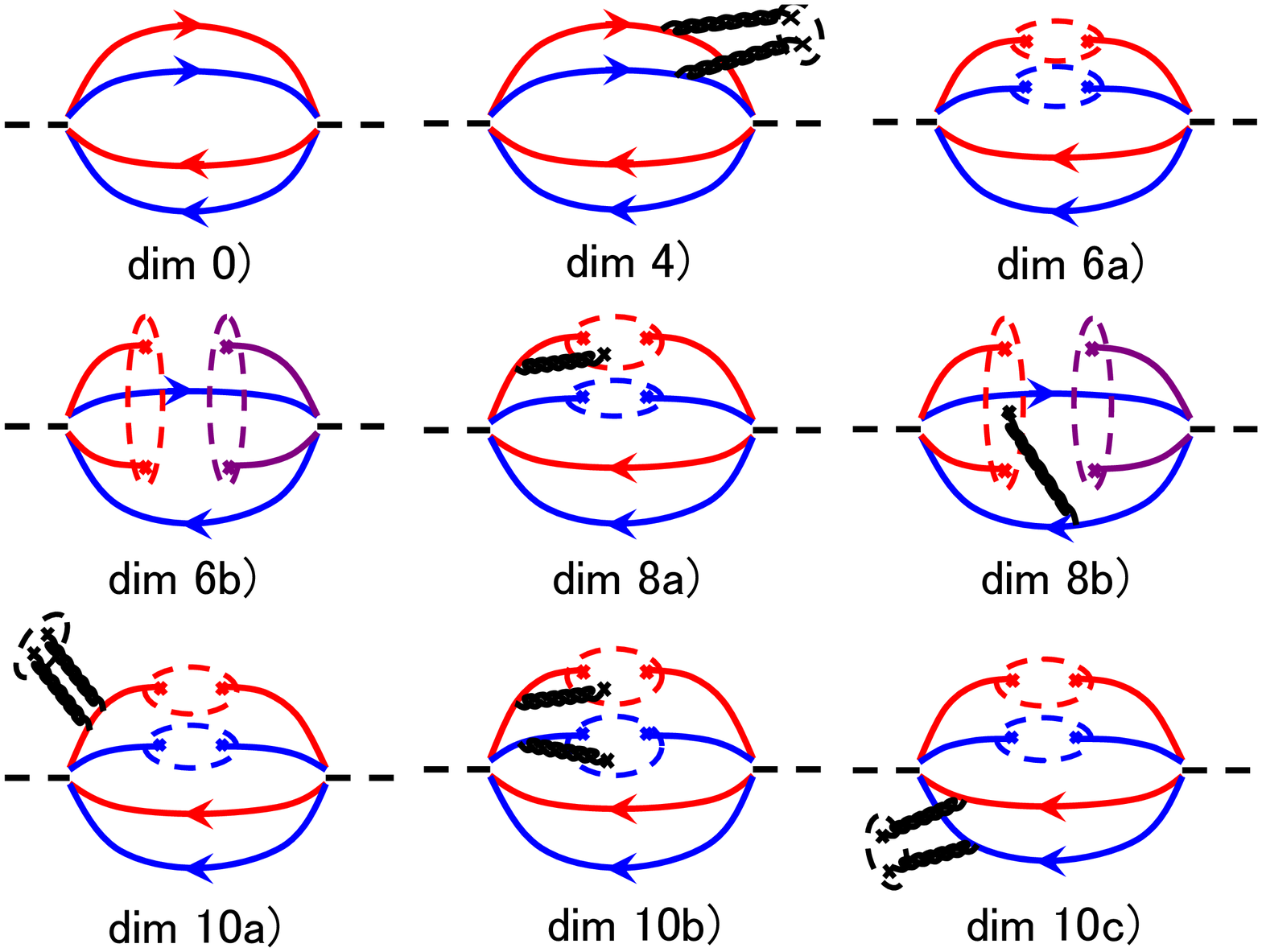}} \\
     \hspace{-0.0cm}
     \vspace{-0.5cm}
     \resizebox{85mm}{!}{\includegraphics{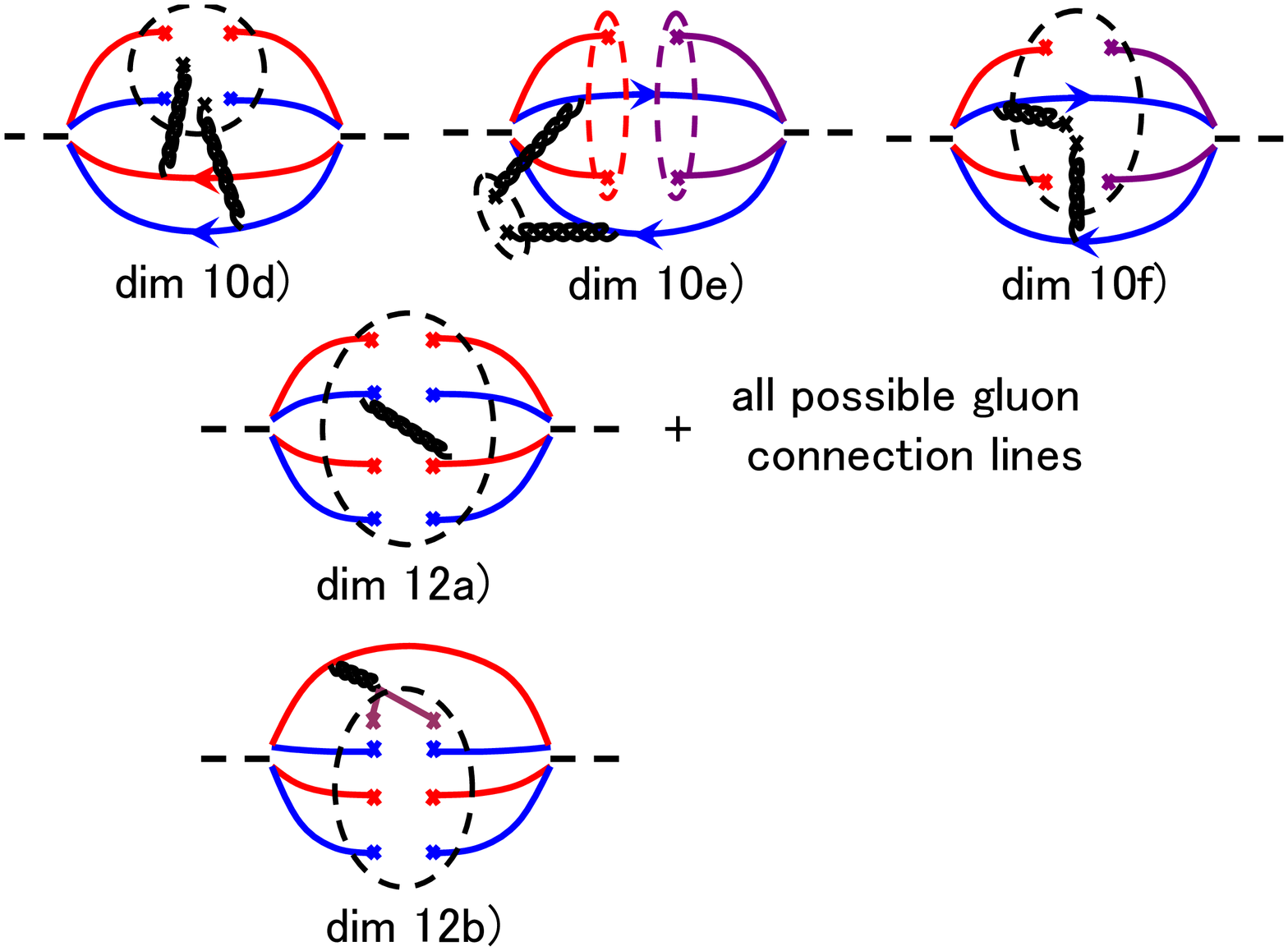}} \\
    \end{tabular}
   \vspace{0.3cm}
   \caption{(Color online) The diagrams for the OPE.}
   \label{diagrams}
  \end{center}
\end{figure}

The OPE of the correlation function 
is expressed by the the coefficients $C_i$:
\begin{eqnarray}
\Pi_i^{ope}(q^2)
= \sum_{j=0}^{4}C_{2j}\ (q^2)^{4-j}\log(-q^2)
+ \sum_{j=1}^\infty\frac{C_{8+2j}}{(q^2)^j}.\,
\label{eq:operesults}
\end{eqnarray}
The relevant diagrams to calculate the OPE are shown in Fig.\ref{diagrams}.

We use
${\rm c}\equiv {\cos}\theta,\ {\rm s} \equiv {\sin}\theta$,
and $\langle \bar{u}u \rangle = \langle \bar{d}d \rangle 
= \langle \bar{s}s \rangle= \langle \bar{q}q \rangle $
for the simplication of the OPE expressions.
We also define $N$ as a number of the annihilation diagrams,
$N=1$ (for octet), $N=2$ (for $\sigma$) and $N=4$ (for singlet).

For the terms from dim.0 to dim.4, we have  
\begin{eqnarray}
\!\!&&\!\! C_0 = - ({\rm c}^2 + {\rm s}^2)\ \frac{1}{2^{12}\ 3\ 5\pi^6},\\ 
&&\!\! C_4 = - ({\rm c}^2 + {\rm s}^2)\ \frac{ \langle \frac{\alpha_s}{\pi}G^2
  \rangle }{2^{10}\ 3\ \pi^4} \times 2,
\end{eqnarray}
where ``$\times ...$'' is the multiplication factor
from the permutation of diagrams.

From dim.6, the annihilation diagrams begin to appear. 
The dim.6 terms are calculated as
\begin{eqnarray}
&&\!\! C_{6a} = ({\rm c}^2 - {\rm s}^2)\ \frac{ \langle \bar{q}q \rangle^2 }
   {2^3\ 3\ \pi^2} \times 2,\\
&&\!\! C_{6b} = - ({\rm c}^2 + {\rm s}^2-2{\rm sc})\ \frac{ \langle \bar{q}q \rangle^2 }
   {2^5\ 3\ \pi^2} \times N, 
\end{eqnarray}
where the suffix of $C$ stands for  the diagram in Fig.\ref{diagrams}.

The dim.8 terms provide most important contributions.
These are the first contributions beyond the polynomial
$s^n(n>0)$ and give the most of the low energy correlation. 
They are calculated as
\begin{eqnarray}
&&\!\! C_{8a} = - ({\rm c}^2 + {\rm s}^2)\ 
 \frac{\langle \bar{q}q \rangle \langle \bar{q} g_s \sigma G q \rangle}
   {2^{4}\ 3\ \pi^2} \times 4,\\
&&\!\! C_{8b} = - ({\rm c}^2 + {\rm s}^2-2{\rm sc})\ 
 \frac{\langle \bar{q}q \rangle \langle \bar{q} g_s \sigma G q \rangle}
   {2^{6}\ 3\ \pi^2} \times 2{N}.
\end{eqnarray}
Note that the annihilation diagrams can give substantial contributions
roughly $\sim 40$\% of the total of dim.8.

\begin{figure*}[floatfix]
 \hspace{0.7cm}     {\large {\bf Effective mass plot} }
 \vspace{0.0cm}\\
 \begin{center}
      \hspace{0.8cm} {\bf singlet} \hspace{3.1cm} {\bf octet} \hspace{2.5cm} {\bf sigma meson}
 \vspace{-0.2cm}
     \resizebox{140mm}{!}{\includegraphics{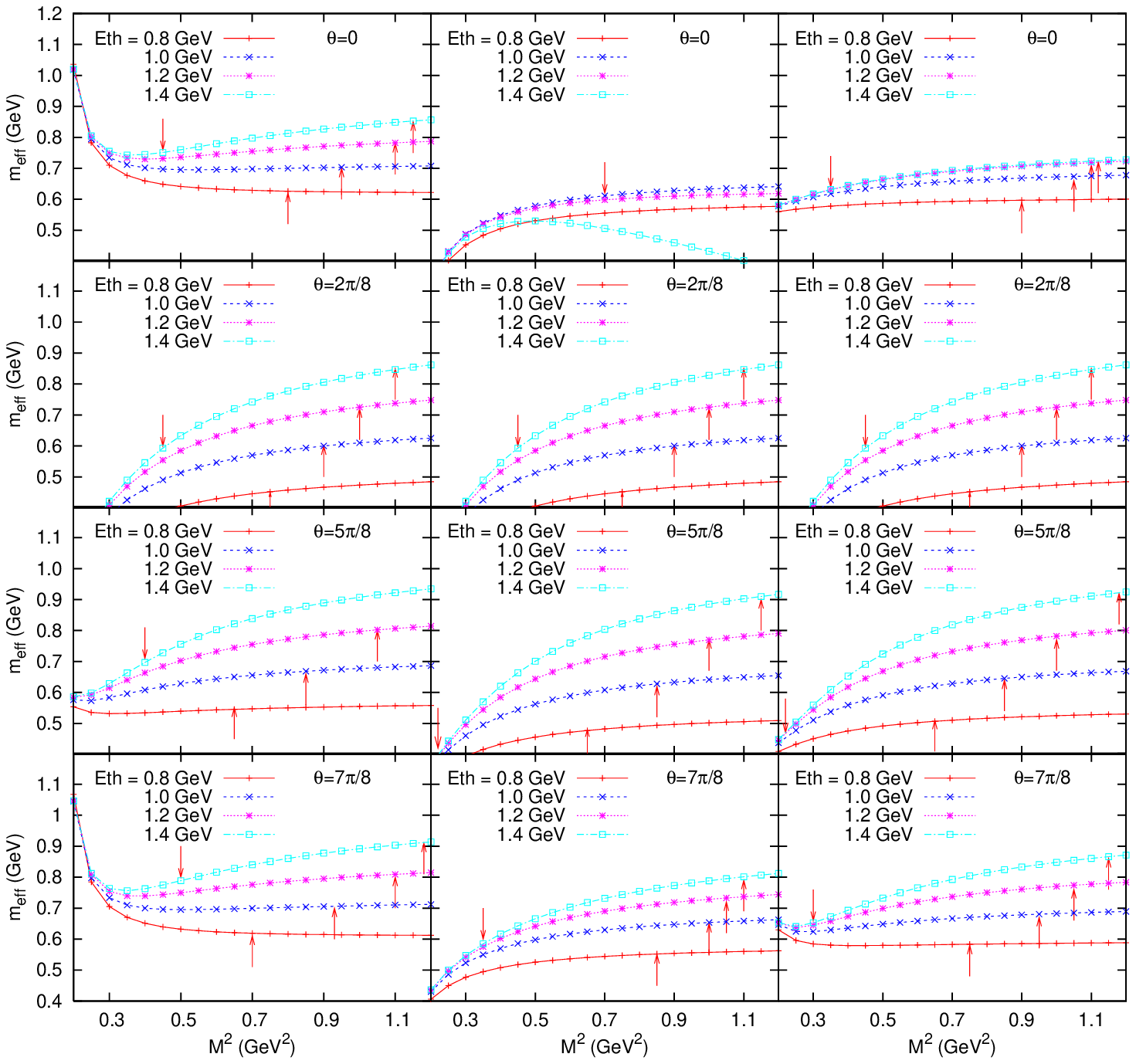}} 
      \vspace{0.5cm} \\
  \caption{(Color online) The mixing angle ($\theta=0,\ 2\pi/8,\ 5\pi/8,\ 7\pi/8$)
  dependence of the effective mass plots for singlet, octet, and $\sigma$
  meson cases. 
  For $\theta=0$ in the octet case,
  we show only the lower bound of the Borel window since
  the spectral condition is not satisfied in the region $E_{th}=0.8\sim
  1.4$ GeV.
  }\label{listmass}
  \vspace{-0.5cm}
\end{center}
\end{figure*}

From dim.10, the OPE begins to converge, but still
substantial contributions give to the correlation function. The coefficients of the
dim.10 terms are obtained as  
\begin{eqnarray}
&&\!\!\!\! C_{10a} = ({\rm c}^2 - {\rm s}^2)\ 
 \frac{ \langle \bar{q}q \rangle^2 \langle \frac{\alpha_s}{\pi}G^2 \rangle}
   {2^3\ 3^3} \times 4,\\
&&\!\!\!\! C_{10b} = ({\rm c}^2 - {\rm s}^2)\ 
 \frac{\langle \bar{q} g_s \sigma G q \rangle^2}
   {2^{6}\ 3\ \pi^2} \times 2,\\
&&\!\!\!\! C_{10c} = ({\rm c}^2 - {\rm s}^2)\ 
 \frac{ \langle \bar{q}q \rangle^2 \langle \frac{\alpha_s}{\pi}G^2 \rangle}
   {2^{4}\ 3^2\ \pi^2} \times 2,\\
&&\!\!\!\! C_{10d} = ({\rm c}^2 - {\rm s}^2)\ 
 \frac{11\langle \bar{q} g_s \sigma G q \rangle^2}
   {2^{10}\ 3^2\ \pi^2} \times 2,\\
&&\!\!\!\! C_{10e} = - ({\rm c}^2 + {\rm s}^2 - 2{\rm sc})\ 
 \frac{ \langle \bar{q}q \rangle^2 \langle \frac{\alpha_s}{\pi}G^2 \rangle}
   {2^{5}\ 3^2} \times N,\\
&&\!\!\!\! C_{10f} = - ({\rm c}^2 + {\rm s}^2 - 2{\rm sc})\ 
 \frac{ \langle \bar{q} g_s \sigma G q\rangle^2 }
   {2^{9}\ 3\ \pi^2} \times {N}.
\end{eqnarray}
In our analysis, we calculate the OPE up to dim.12,
because further dimension terms have 
no large enhancement factor of $\sim (4\pi)^2$ coming
from cutting quark loops \cite{KHJ}. 
There are many diagrams for the dim.12 terms
because we have many ways to attach the gluon line to
the quark lines.
The results after adding the all possible diagrams are
\begin{eqnarray}
&&\!\!\!\!\!\! C_{12a} = \frac{\pi \alpha_s \langle \bar{q}q \rangle^4}{3^2} \\ 
&&\!\!\!\!\!\! \times [ 8 ({\rm c}^2 + {\rm s}^2) - 3N ({\rm c}^2 + {\rm s}^2 -
 2{\rm sc}) ], \\
&&\!\!\!\!\!\! C_{12b} = \frac{2^4 \pi \alpha_s \langle \bar{q}q \rangle^4}{3^5}
 [ 2 ({\rm c}^2 - {\rm s}^2) - N ( {\rm c}^2 - {\rm sc}) ].
\end{eqnarray}
There are no terms having the structure other than the above, such as
$\langle \bar{q} q\rangle \langle \bar{q} g_s \sigma G q \rangle \langle \frac{\alpha_s}{\pi}G^2 \rangle $.

\begin{figure*}[floatfix]
 \hspace{0.7cm}         {\large {\bf Effective residue plot} }
 \vspace{0.0cm}\\
 \begin{center}
      \hspace{0.8cm} {\bf singlet} \hspace{3.1cm} {\bf octet} \hspace{2.5cm} {\bf sigma meson}
 \vspace{-0.2cm}
     \resizebox{140mm}{!}{\includegraphics{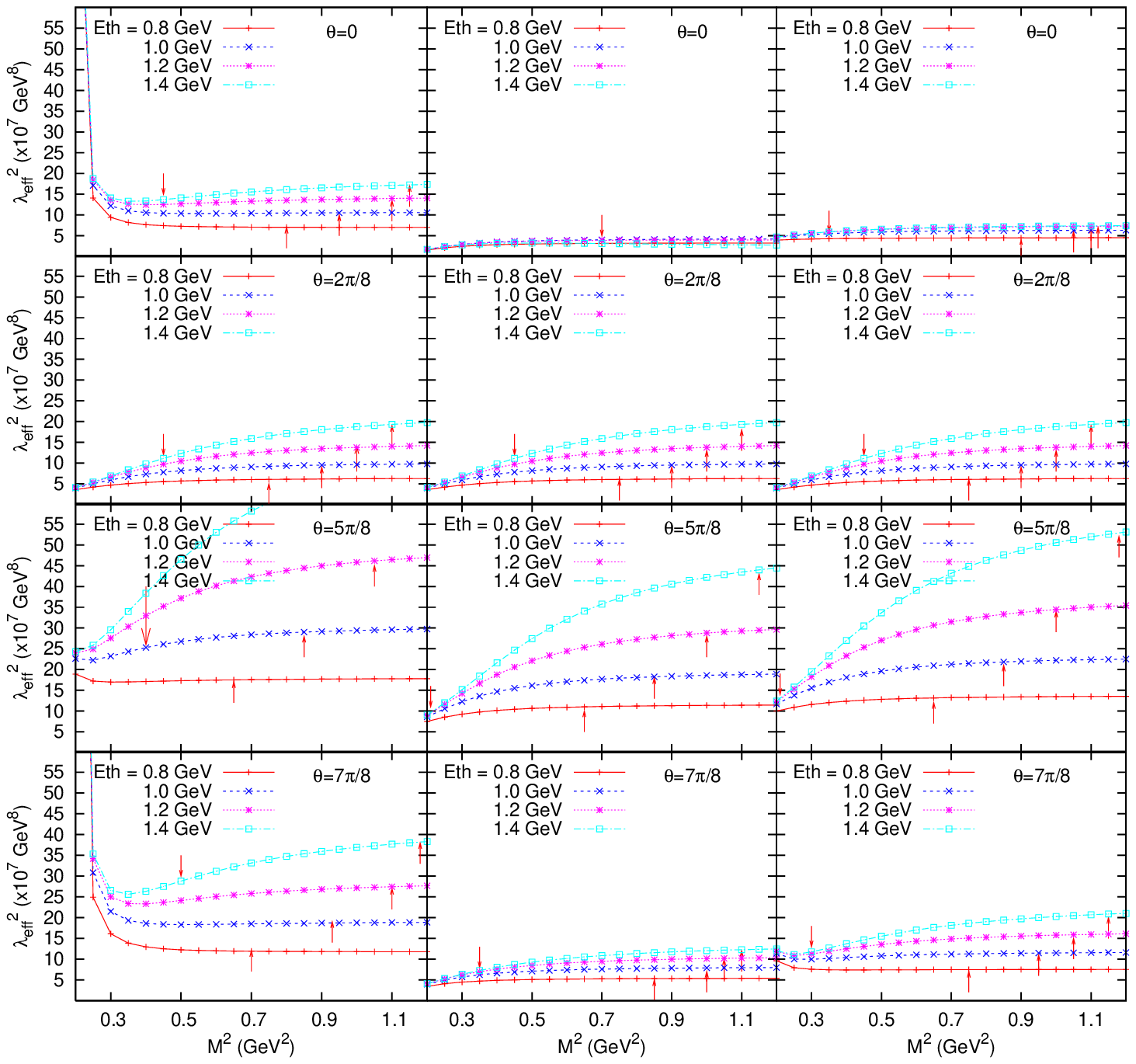}} 
  \vspace{0.5cm} \\
  \caption{(Color online) The mixing angle ($\theta=0,\ 2\pi/8,\ 5\pi/8,\ 7\pi/8$) 
  dependence of the effective residue plots for singlet, octet, and
  $\sigma$ meson cases.
  For $\theta=0$ in the octet case,
  we show only the lower bound of the Borel window since
  the spectral condition is not satisfied in the region $E_{th}=0.8\sim
  1.4$ GeV.}\label{listresi}
  \vspace{-0.5cm}
\end{center}
\end{figure*}

\section{$\theta$ dependence of the effective residue and mass}
\label{sec:thetadep}

In this Appendix, we discuss the detail of the
$\theta$ dependence.
We would like to draw a rough sketch of the allowed region for the 
mixing angle to achieve wider Borel window and better pole isolation
following to the discussion in Sec.\ref{sec:IF}. 

We show in Fig.\ref{listmass} 
the effective mass obtained by the QSR
as functions of the mixing angle $\theta$ $(0,\ 2\pi/8,\ 5\pi/8,\ 7\pi/8)$
and the Borel mass $M^{2}$.
The threshold is taken as $E_{th}=0.8,\ 1.0,\ 1.2,\ 1.4$ GeV in the
same way as subsection \ref{sec:SingletOctet} and \ref{sec:sigmaSR}.
We also show the effective residue in Fig.\ref{listresi} 
with the same $\theta$ and $E_{th}$ as the effective mass plots.

The essential points were already mentioned in subsection \ref{sec:SingletOctet}.
Here we give more detailed explanations.
\begin{itemize}
  \item the Borel windows can be established in all the mixing angles 
	except $\theta \sim 0$ for the flavor octet case.
	For $\theta \sim 0$, the spectral condition 
	${\rm Im}\Pi^{ope}(s)\ge 0$ is slightly violated 
	unless we take $E_{th}$ as large as
	$\sim 2.0\ {\rm GeV}$.
	Thus the region around $\theta =0$ was avoided in subsection
	\ref{sec:SingletOctet} to compare singlet and octet
	cases in the region where sum rules for both cases well
	satisfy the criteria in subsection \ref{sec:IF} .
  \item the results for the physical quantities are not very sensitive to 
	the choice of the mixing 
	angle $\theta$ in the  
	regions of $\theta =0 \sim 2\pi/8$ and $6\pi/8 \sim \pi$
	both in the flavor singlet and octet cases.
	This is most clearly reflected in the effective
	residue plots in Fig.\ref{listresi}.
	On the other hand, except these regions,
	the effective residua strongly depend on the mixing angle and the threshold.
	We interpret that these strong dependences 
	are mainly related to contributions from scattering states.
	This is because the residua in these $\theta$ regions show the large $E_{th}$
	dependence, which indicates
	that the resonance pole is not isolated enough. 
  \item one of the best value of the mixing angle both for the singlet
	and octet is $\theta \simeq 7\pi/8$.
	If we consider only the singlet and $\sigma$ meson cases,
	the best result is obtained with $\theta=0$.
	In this case, the threshold dependence is extremely small,
	and we can expect that there are small contaminations
	in the region from resonance pole to $E_{th}$, i.e.,
	the resonance is isolated. 
  \item the residua of the singlet states have larger values than those
	of the octet states in almost all region of $\theta$.
  \item the effective masses of the octet case are typically smaller
	than those of the singlet, and the Borel stability is worse in the octet case. 
\end{itemize}

In addition to these, we put one remark 
on the chiral representation of the interpolating fields:
in the case of $J(\pi/4)$, i.e.,
the combination of $[({\bf 8},{\bf 1}) \otimes ({\bf 1}, {\bf 8})]$ 
and $({\bf 1},\bar{\bf 1})$ representations,
the contributions from annihilation diagrams vanish
(see also OPE expression),
and the results of singlet, octet, and $\sigma$ meson
degenerate.

Taking into account the above all remarks,
we evaluate the mass for singlet, octet, and $\sigma$
meson cases.
We mainly consider the cases of $\theta=0\ {\rm and}\ 7\pi/8$ in Fig.\ref{listmass}.
For the singlet case, 
the typical effective mass ranges are
from $600\ {\rm MeV}$ to $850\ {\rm MeV}$ in wide Borel windows
with reasonable Borel stability.
The $E_{th}$ dependence is also reasonalbly small.
For the octet case, we must avoid $\theta\sim0$.
The $\theta\sim 7\pi/8$ case was already discussed
in Sec.\ref{sec:SingletOctet}.
Finally, for $\sigma$ meson case,
the behavior of effective mass plot is better than both singlet
and octet cases.
We evaluate the mass as $600\sim800$ MeV.

\end{document}